\documentclass[reprint, amsmath,amssymb, aps,onecolumn]{revtex4-2}

\usepackage{graphicx}
\usepackage{subcaption}
\graphicspath{{./}}
\usepackage{bm}
\usepackage{mathrsfs}  
\usepackage{tikz,xcolor,hyperref}
\usepackage{ragged2e}
\usepackage{cancel}
\usepackage[normalem]{ulem}

\definecolor{lime}{HTML}{A6CE39}
\DeclareRobustCommand{\orcidicon}{
	\begin{tikzpicture}
	\draw[lime, fill=lime] (0,0) 
	circle [radius=0.16] 
	node[white] {{\fontfamily{qag}\selectfont \tiny ID}};
	\draw[white, fill=white] (-0.0625,0.095) 
	circle [radius=0.07];
	\end{tikzpicture}
	\hspace{-2mm}
}
	
\foreach \x in {A, ..., Z}{\expandafter\xdef\csname orcid\x\endcsname{\noexpand\href{https://orcid.org/\csname orcidauthor\x\endcsname}
			{\noexpand\orcidicon}}
}

\begin{document}

\preprint{APS/123-QED}

\title{Reconfiguration and dynamics of clamped fibers under finite-amplitude surface gravity waves}

\author{Giulio Foggi Rota \orcidA{}}
\author{Alessandro Chiarini \orcidB{}}\altaffiliation{Present address: Dipartimento di Scienze e Tecnologie Aerospaziali, Politecnico di Milano, via La Masa 34, 20156 Milano, Italy}
\author{Marco Edoardo Rosti \orcidC{}}
\email[E-mail for correspondence: ]{marco.rosti@oist.jp}
\affiliation{Complex Fluids and Flows Unit, Okinawa Institute of Science and Technology Graduate University, 1919-1 Tancha, Onna-son, Okinawa 904-0495, Japan.}

\begin{abstract}
\justifying
We investigate the behaviour of a flexible stem completely submerged under a surface gravity wave of finite amplitude using fully resolved direct numerical simulations. By varying the rigidity of the stem over ten orders of magnitude, we explore its motion in the drag-dominated regime with realistic air and water properties. Our findings reveal two distinct structural responses of the stem depending on the ratio between its natural frequency ($f_{nat}$) and the wave frequency ($f_{wave}$).
For $f_{nat}/f_{wave} \gg 1$, the stem maintains on average a straight configuration and exhibits streamwise oscillations in phase-opposition with the wave, moving symmetrically with respect to the vertical direction.
Conversely, for $f_{nat}/f_{wave} \ll 1$, the stem reconfigures under the influence of the Stokes drift, bending forward and breaking the symmetry, and exhibits oscillations that are more coherent with the surrounding flow field. Resonance is observed when $f_{nat} \approx f_{wave}$.
These results provide new insights into the dynamics of slender vegetation and man-made structures in wave fields, offering valuable implications for marine biology and engineering.
\end{abstract}

\maketitle

\section{Introduction}

Flexible structures anchored in moving fluids can significantly alter their shape compared to their configuration at rest. This \textit{reconfiguration} \cite{alben-shelley-zhang-2002} is a common defensive behaviour in the botanical world, as changes in shape often mitigate aerodynamic loads on exposed surfaces \cite{gosselin-2019}. Tree leaves, for instance, are typically flexible and compliant to atmospheric winds rather than rigidly opposing to their pressure. Similarly, marine vegetation often features slender morphologies that bend easily under surface waves and adapt to varying flow conditions \cite{luhar-nepf-2011}.
Reconfiguration is also widely studied and employed in engineering applications \cite{alben-2008, ifju-albertani-shyy-2008, gosselin-langre-machadoalmeida-2010}, where the deformation of artefacts interacting with marine and atmospheric flows is carefully planned to minimise operational disruptions (e.g., suspension bridges) and maximise efficiency (e.g., airplane wings). The deformation of flexible structures under hydraulic and aerodynamic loads can be static \cite{luhar-nepf-2011} or dynamic \cite{bearman-1984, rosti-etal-2018}, depending on multiple structural and environmental factors. Typically, when the flow in which the structure is submerged exhibits significant temporal variations, the structural response is also unsteady and may reveal its \textit{natural dynamics} in multiple forms. This is evident in the case of a flexible stem submerged in a surface gravity wave, which is indeed the focus of the present numerical investigation.

Few studies have investigated the dynamics of a single flexible stem anchored in a moving fluid over the last years.
Starting from the seminal experiments in soap films \cite{zhang-etal-2000,jung-etal-2006}, drag models have been developed \cite{maza-lara-losada-2013, luhar-nepf-2016}, and the structural response of flexible objects to different kinds of fluid forcing has been assessed \cite{bagheri-mazzino-bottaro-2012,lacis-etal-2014,luhar-nepf-2016, jacobsen-etal-2019}. 
In turbulence, \citet{foggirota-etal-2024} numerically investigated the motion of filamentous marine vegetation (similar to \textit{Zostera marina}), highlighting how the time-averaged shape and the response of a single stem to the turbulent fluctuations change with its structural rigidity.
The case of submerged vegetation clamped under surface gravity waves \cite{lighthill-2001,landau-lifshitz-2003,chabchoub-akhmediev-hoffmann-2012,devita-verzicco-iafrati-2018,devita-etal-2021,lorenzo-etal-2023,mcwilliams-2023} has received particular attention due to its environmental and biological relevance. 
\citet{leclercq-delangre-2018} addressed the reconfiguration of an elastic blade with fixed structural properties in an oscillatory flow, identifying different kinematic regimes varying the frequency and the amplitude of the incoming wave. 
\citet{jacobsen-etal-2019} considered a flexible blade and described the fluid-structure interaction in terms of displacements, phase lags, relative velocities and force coefficients.
\citet{mullarney-henderson-2010} and \citet{kumar-etal-2021} modelled and experimentally investigated the motion of a flexible stem in the \textit{drag-dominated} regime, where drag forces dominate over fluid inertia.   
In this case, representative of multiple scenarios of practical interest, the reconfiguration is limited and the stem typically sways with the period of the incoming waves without exhibiting an evident resonant behaviour \cite{kumar-etal-2021}.
However, the experimental nature of most studies \cite{mullarney-henderson-2010, jacobsen-etal-2019, kumar-etal-2021} has prevented an extensive sampling of the structural parameters space, favouring instead the investigation of multiple wave amplitudes and frequencies. 
It is thus still unclear whether and to what extent the natural dynamics of the stem influences its motion under the incoming waves. 

In this study we investigate the drag-dominated motion of a filamentous marine plant comprising a single flexible stem fully submerged in water under a monochromatic surface gravity wave of finite amplitude. We perform fully-resolved and coupled simulations of the multiphase system (air, water, and structure) and examine how the oscillatory motion of the stem changes with its rigidity, which is indeed the parameter controlling its natural dynamics. We explore how the swaying trajectory of the stem varies with its rigidity/natural frequency and observe a small but non-negligible reconfiguration in the most flexible cases. In these cases, indeed, due to the Stokes drift \cite{stokes-1847,longuett-stoneley-1953,santamaria-etal-2013,bremer-breivik-2017} the stem exhibits a small deflection in the streamwise direction due to the momentum exchange with the net mass flux under the water surface. Notably, when the natural frequency of the stem matches that of the incoming wave, we observe a resonance effect 
that has never been detected so far in the drag-dominated regime. 

The remaining part of the paper is organised as follows. In \S\ref{sec:setup_methods}, we introduce the setup of our numerical experiments and describe the methods adopted to address the dynamics of the fluids (air and water) and of the structure, as well as their mutual interaction. In \S\ref{sec:results}, we delve into the dynamics of the stem by sampling the trajectory described by its tip over time. We elucidate the dependence of such trajectory on the rigidity of the stem, relate it to the Stokes drift and quantify the response of the stem in terms of gain and phase. Additionally, we investigate the resonance observed when the natural frequency of the stem matches that of the incoming surface wave. Eventually, in \S\ref{sec:conclusions} we summarise the main outcomes of the investigation.

\section{Setup and methods}
\label{sec:setup_methods}

\begin{figure}
\begin{minipage}{0.43\textwidth}
        \vspace{-250pt} 
        \includegraphics[width=\textwidth]{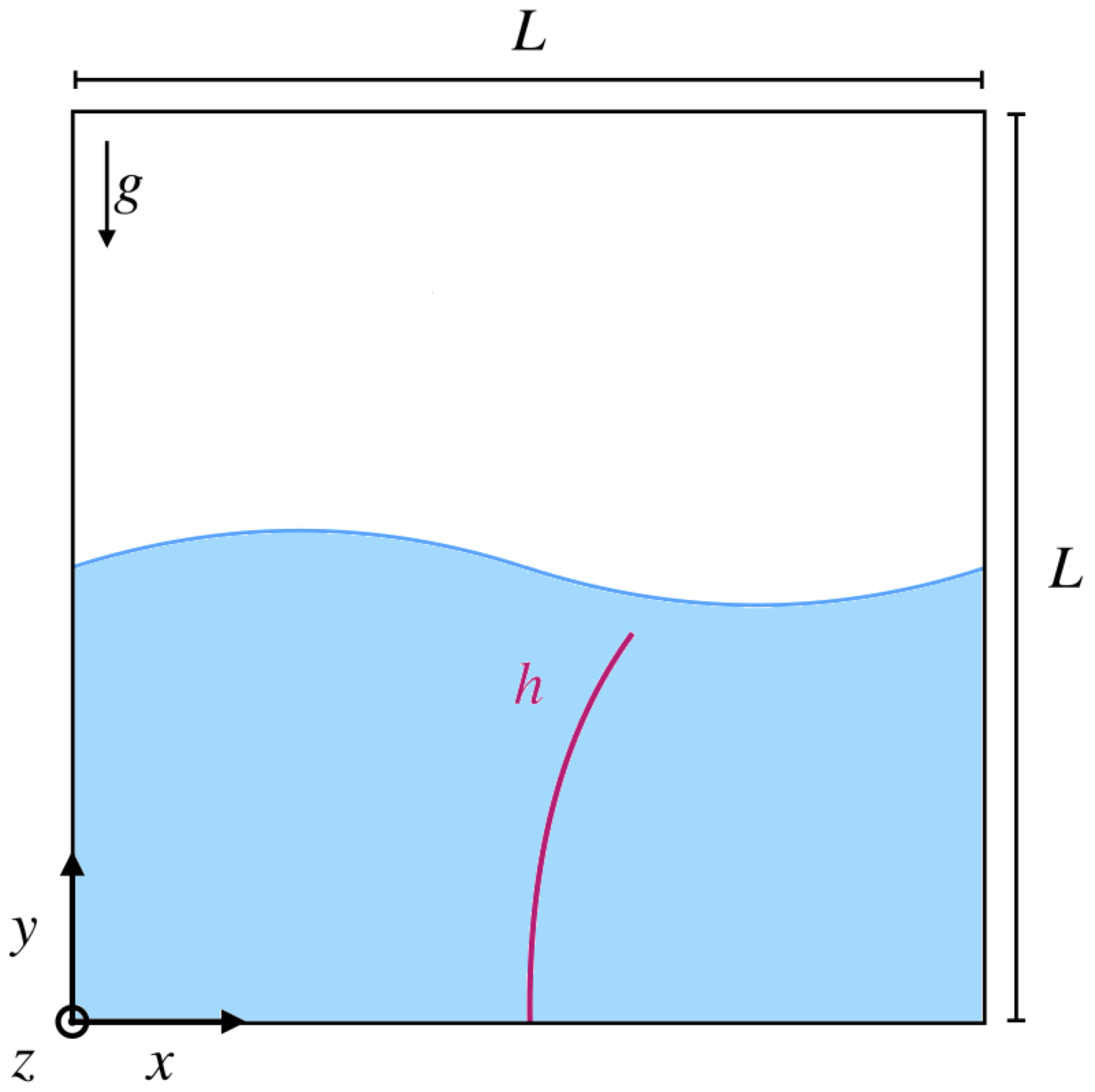}
\end{minipage}%
	\includegraphics[width=0.48\textwidth]{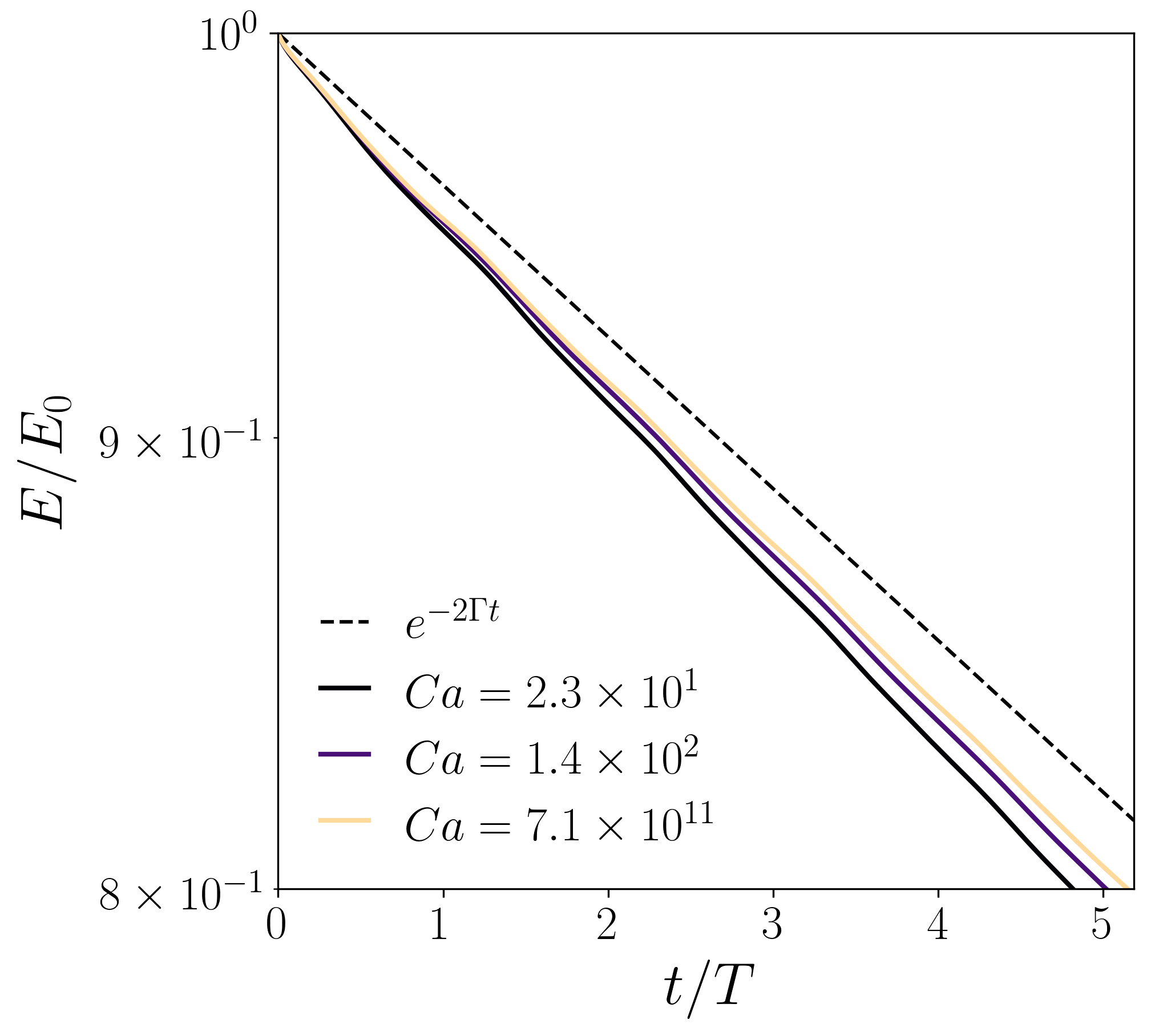}
	\put(-470,210){a)}
	\put(-220,210){b)}
	\caption{In panel $a$ we show a bi-dimensional sketch of our computational domain, sliced in the middle. Panel $b$, instead, reports the time decay of the mechanical energy of the water measured in our simulations (continuous lines) against the analytical prediction of \citet{landau-lifshitz-2003} (dashed line). Darker lines denote the presence of stems of increasing rigidity, in the range $Ca\in[2.3 \times 10^{1}, 7.1 \times 10^{11}]$.}
	\label{fig:sketch}
\end{figure}

We investigate by direct numerical simulations the dynamics of a flexible stem of length $h$ fully submerged in water under a monocromatic surface gravity wave of wavelength $L$ and finite amplitude $A$. We perform the simulations in a three-dimensional computational box with size $L$ along the longitudinal $x$ direction (aligned with the wave propagation) and the vertical $y$ direction, and size $L/4$ along the transversal (spanwise) $z$ direction. The box is half-filled with water (up to $y/L=0.5$), and it lays in a gravitational field $\mathbf{g}=(0,-g,0)$ oriented with the negative $y$ direction. The slender flexible stem is vertically clamped at the centre of the bottom surface, as pictorially represented in panel $a$ of Fig.~\ref{fig:sketch}. 
The top and bottom faces of the box are solid walls at which no fluid slip nor penetration are allowed, while periodicity is enforced at all the other faces.
We have verified that the imposition of free slip boundary conditions at the top and bottom faces does not produce any appreciable effect on the wave field, nor does a variation of the domain extent in the vertical direction, as expected in the case of deep water waves \citep{landau-lifshitz-2003}.

The fluid motion within the domain is governed by the incompressible Navier--Stokes equations, i.e.
\begin{eqnarray}
   &\nabla \cdot \mathbf{u} = 0,
   \label{eq:mass}\\
   &\displaystyle \frac{\partial \mathbf{u}}{\partial t} + \nabla \cdot (\mathbf{u}  \mathbf{u})= \frac{1}{\rho_f}\left(- \nabla p + \nabla \cdot \boldsymbol{\tau}\right) + \mathbf{f}.
   \label{eq:momentum}
\end{eqnarray}
Here $\mathbf{u}$ and $p$ denote the velocity and pressure fields, $\rho _f$ is the local fluid density (being $\rho_w$ in the water and $\rho_a$ in the air), $\boldsymbol{\tau}$ is the stress tensor, and $\mathbf{f}$ accounts for the volume forces.
Since we consider the flow of two immiscible fluid phases (water and air) separated by an interface, the flow velocity is required to be continuous at the interface, while the stress tensor jumps due to the action of the surface tension.
Here, we treat the surface tension as a volume force $\sigma k \delta_s \mathbf{n}$ \cite{popinet-2018}, where $\sigma$ is the surface tension itself, $k$ is the local curvature of the interface, $\mathbf{n}$ its normal unit vector and $\delta_s$ a Dirac $\delta$-function that is nonzero only on the interface.
Furthermore, we introduce a body force $\mathbf{f}_{\mathbf{\rm{IBM}}}$ to deal with the presence of the immersed object.
Accounting for all this, in Eqn.~\eqref{eq:momentum} $\mathbf{f}$ writes $\mathbf{f}=\mathbf{f}_{\mathbf{\rm{IBM}}}+\mathbf{g}+\sigma k \delta_s \mathbf{n}$.
The  motion of the flexible stem is described by an extended version of the distributed-Lagrange-multiplier/fictitious-domain (DLM/FD) formulation of the continuum equations \cite{yu-2005}, which represents a generalisation of the Euler-Bernoulli beam model that allows for finite deflections and retains the inextensibility constraint \cite{banaei-rosti-brandt-2020}.
Indicating with $\mathbf{X}(s,t)$ the position of a point on the neutral axis of the stem as a function of the curvilinear abscissa $s$ and time $t$, and introducing the linear density difference between the stem and the water $ \Delta \Tilde{\rho} = (\rho_s-\rho_w)\pi r^2 $ with $r$ denoting the cross-sectional radius of the stem, its structural dynamics is described by 
\begin{eqnarray}
    &\Delta \Tilde{\rho} \displaystyle \frac{\partial^2 \mathbf{X}}{\partial t^2} = 
    \displaystyle \frac{\partial}{\partial s}\left(T\displaystyle \frac{\partial \mathbf{X}}{\partial s}\right) - \gamma \displaystyle \frac{\partial^4 \mathbf{X}}{\partial s^4} - \mathbf{F}, 
    \label{eq:eulerBernoulli}\\
    &\displaystyle \frac{\partial \mathbf{X}}{\partial s} \cdot \displaystyle \frac{\partial \mathbf{X}}{\partial s} = 1,
    \label{eq:inextensibility}
\end{eqnarray}
where $T$ is the tension enforcing inextensibility and $\mathbf{F}$ is the force acting on the stem computed by the Lagrangian IBM to couple them with the fluid.
The structural model adopted assumes a slender structure where the internal dissipation of energy is negligible. As such, no explicit dissipative term appears in the structural equations, and energy is only dissipated by the fluid viscosity. We complement Eqns. \eqref{eq:eulerBernoulli} and \eqref{eq:inextensibility} with an appropriate set of boundary conditions, imposing $\mathbf{X}\rvert_{s=0}=\mathbf{X_0}$ along with ${\partial\mathbf{X}}/{\partial s}\rvert_{s=0}=(0,1,0)$ at the clamp and ${\partial^3\mathbf{X}}/{\partial s^3}\rvert_{s=h}={\partial^2\mathbf{X}}/{\partial s^2}\rvert_{s=h}=\mathbf{0}$ along with $T\rvert_{s=h}=0$ at the free end.

The simulations are initialised with the fluid at rest and the flexible stem being in a perfectly vertical configuration, i.e. still and undeflected under the surface. We then drive the fluid with an unsteady volume forcing with an amplitude that grows linearly in time, such that after $3T$ ($T$ is the wave period; see the following) the fluid velocity field matches that of the analytical inviscid solution for a gravity wave of amplitude $A$ and wavelength $L$ \cite{lighthill-2001}, i.e.
\begin{equation}
\label{eq:fluidFun}
	\mathbf{u_0}(x,y; t)=
	\left\{
	\begin{array}{ll}
             u_0 \\
	v_0 \\
	w_0
            \end{array} 
            \right\}
            =
	\left\{
	\begin{array}{ll}
             \phi \frac{A}{L\kappa} \omega e^{\kappa (y- \frac{L}{2})} cos(\kappa x -\omega t) - (1-\phi) \frac{A}{L\kappa} \omega e^{\kappa (\frac{L}{2}-y)} cos(\kappa x -\omega t)\\
	\phi \frac{A}{L\kappa} \omega e^{\kappa (y- \frac{L}{2})} sin(\kappa x -\omega t) + (1-\phi) \frac{A}{L\kappa} \omega e^{\kappa (\frac{L}{2}-y)} sin(\kappa x -\omega t)\\
	0
            \end{array} 
            \right\}
\end{equation}
where $\kappa = {2\pi}/{L}$ is the wavenumber ($\kappa$ constitutes the only non-zero component of the wave vector, aligned with the $x$ direction) and $\omega=\sqrt{g \kappa}$ is the angular frequency, which relates to the wave period $T$ through the relation $T=2\pi/\omega$. 
Here $\phi$ is the volume-of-fluid function, which is $1$ in water, $0$ in air and $\phi \in (0,1)$ at the interface.
Such velocity field is compatible with the following perturbation of the air/water surface $y_s$, i.e.
\begin{equation}
\label{eq:surfFun}
	y_{s} (x,t)= \frac{L}{2} + \frac{A}{L\kappa} cos(\kappa x -\omega t).
\end{equation}

We then maintain the forcing constant for an additional wave period to allow the flow to settle, and eventually release the wave (i.e. we set the volume forcing to zero) at $t=0$. We thus perform our measurements for $t \ge0 $ when the system is free to evolve, by solving the whole body of the multiphase Navier-Stokes equations to capture the back-reaction of the structure on the fluid and ensure realistic results.
This procedure allows for a gradual adaptation of the flexible stem to the fluid motion, and ensures compatible velocities for the fluid and the structure at the beginning of the measurements.
Such compatibility would not be attained simply placing a straight and still stem in a fully developed wave field.  
We have also tested a different analytical form of the initial velocity field, accounting for the finite depth of the fluid \cite{lighthill-2001}, without noticing any appreciable variation in the results. 
Since we do not impose any external forcing while performing our measurements and we work with realistic fluids, the wave decays in time for $t \ge 0$ until it is completely dissipated by the fluid viscosity. 
The time scale of the wave decay is anyway significantly larger than that of the stem motion, which thus sees a steady travelling wave of approximately constant amplitude over all the time periods considered in this study. Further details are discussed in the final part of this section. 

The behaviour of the multiphase system at hand depends on a number of parameters and dimensionless groups. We set realistic water-to-air density and dynamic viscosity ratios, i.e. $ {\rho_w}/{\rho_a} = 800$ and $ {\mu_w}/{\mu_a} = 55$ respectively. 
We consider a finite wave of amplitude $A/L=0.1$ (at $t=0$) that evolves without breaking at the interface between the two immiscible fluid phases. According to \citet{digiorgio-pirozzoli-iafrati-2022}, indeed, finite-amplitude waves break for $A/L\gtrapprox 0.33$.
The Reynolds number of the water flow based on the wavelength and phase speed \cite{devita-etal-2021} (quantifying the relative importance of inertial effects over viscous effects) is set to $Re_w=\rho_w\sqrt{g L^3}/\mu_w=10000$, while that of the air flow follows from the previously chosen parameters as $Re_a=\rho_a\sqrt{g L^3}/\mu_a=687.5$. 
After testing a realistic value of $\sigma$, corresponding to a Weber number $We = \sqrt{({\rho_w g L^2})/{\sigma}}\approx 110$ \cite{iafrati-2009,digiorgio-pirozzoli-iafrati-2022}, we compare the stem motion with that attained for a much higher $We$ without noticing any appreciable difference.
We thus observe that the interface dynamics does not play any dominant role in the problem at hand, and simplify it by working in the limit $We\rightarrow \infty$.
We thus set an arbitrarily small value of $\sigma$, corresponding to $We \approx 3500$, and employ it throughout all the simulations.
Multiple buoyant flexible stems of length $h/L = 0.45$, density $\rho_s/\rho_w\approx0.95$ and cross-sectional radius $r/L\approx7.8 \times 10^{-3}$ are considered throughout our simulations, varying their bending rigidity $\gamma$ within ten orders of magnitude. We thus span values of the Cauchy number \cite{luhar-nepf-2011} (representing the ratio between the deforming force exerted by the fluid and the elastic restoring force opposed by the stems) in the range $Ca={(\rho_w r h^3 g L)}/{\gamma}\in[2.3 \times 10^{1}, 7.1 \times 10^{11}]$. We also consider different densities of the stem in $\rho_s/\rho_w\in\{0.5,0.75\}$ for the highest and lowest values of $Ca$, to assess the effect of buoyancy on the observed dynamics.

The geometry of the problem, along with the initial condition, set the value of the Keulegan–Carpenter number, $KC$ \cite{keulegan-carpenter-1956}, which represents the relative importance of drag over inertial forces in the stem motion.  
$KC$ can be defined as the ratio between the amplitude of the flow velocity fluctuations and a characteristic length scale of the stem (i.e. its radius); we attain $KC=A L / r \approx 10$. The oscillations of our stem are therefore dominated by drag ($KC > 1$) throughout the whole study, in close similarity to the available recent experimental studies \cite{mullarney-henderson-2010,kumar-etal-2021}. 

In this work we tackle Eqns.~\eqref{eq:mass} and \eqref{eq:momentum} numerically by means of our well tested solver \href{https://www.oist.jp/research/research-units/cffu/fujin}{\textit{Fujin}}. 
Space derivatives are discretised with a second-order central finite difference scheme and the solution is advanced in time with a second-order Adams-Bashforth method.
We resort to a projection-correction algorithm \cite{kim-moin-1985}, solving the Poisson equation for the pressure with an efficient decomposition library (\textit{2decomp}) coupled to an in-place spectral solver based on the Fourier’s series method described by \citet{dorr-1970}.
The flow variables are sampled on a staggered Cartesian grid made by $256$ points homogeneously distributed along the longitudinal $x$ and vertical $y$ directions and $64$ points homogeneously distributed along the spanwise $z$ direction.
The extreme cases of our study (i.e. $Ca=2.3\times 10^1$ and $Ca=7.1\times 10^{11}$) have been replicated for validation purposes by doubling the grid resolution ($512\times512\times128$); the outcome is reported in  Figs.~\ref{fig:response} and \ref{fig:elEng} with empty marks. The good agreement with the results at lower resolution confirms the adequacy of the grid employed for the bulk of our investigation. 
The fluid and the structure are coupled at their interface through a no-slip boundary condition, guaranteed applying the force distribution computed with a Lagrangian immersed boundary method (IBM) \cite{peskin-2002, huang-etal-2007, olivieri-etal-2020-2}.
We employ $100$ IBM points homogeneously distributed along the stem throughout our study, increasing them to $200$ in the validation cases.
Further details on the numerical schemes adopted can be found in the Supplemental Material of our former study \cite{foggirota-etal-2024}.
The structural solver is the same employed in multiple previous investigations \cite{monti-olivieri-rosti-2023, foggirota-etal-2024}, where validation details are also available \cite{foggirota-etal-2024-2}. 
Eqns. \eqref{eq:eulerBernoulli} and \eqref{eq:inextensibility}) are solved as in \citet{huang-etal-2007}, nevertheless here the bending term is treated implicitly \cite{banaei-rosti-brandt-2020} to allow for a larger time step.
To numerically solve the multiphase system at hand we use the volume-of-fluid (VOF) method \cite{hirt-nichols-1981}, where only one set of equations is solved over the whole domain. This is achieved by introducing a velocity vector field $\mathbf{u}$ valid everywhere, found by applying the volume averaging procedure \cite{quintard-whitaker-1994a}
\begin{equation}
	\mathbf{u}=(1-\phi)\mathbf{u_a}+\phi \mathbf{u_w}
\end{equation}
where $\phi$ is the volume-of-fluid function introduced previously. In particular, the isoline at $\phi=1/2$ represents the air-water interface. 
The stress tensor $\tau$ is also written in a mixture form, similarly to the velocity field above.
To close the full system of Eqns.~\eqref{eq:mass} and \eqref{eq:momentum}, one additional transport equation is needed for the VOF function $\phi$,
\begin{equation}
	\displaystyle \frac{\partial \phi}{\partial t}+\nabla \cdot (\mathbf{u}\mathscr{H})=\phi \nabla \cdot \mathbf{u},
	\label{eq:VOF}
\end{equation}
where $\mathscr{H}$ is a colour function equal to $1$ in water and $0$ in air, related to the numerical VOF function $\phi$ through a cell-average procedure.
The formulation of the VOF method adopted here is that introduced by \citet{li-etal-2012} (multi-dimensional tangent of hyperbola for interface capturing - MTHINC - method), which has already been validated against analytical solutions \cite{rosti-devita-brandt-2019} and employed in multiple investigations by us and other authors \cite{devita-etal-2021,hori-etal-2023,cannon-soligo-rosti-2024}. 

To assess the time evolution of the flow field, which sets the unsteady forcing acting on the stem, we compare the viscous damping of the gravity wave in our simulations with that predicted by \citet{landau-lifshitz-2003}. 
According to their arguments, the time evolution of the total wave energy $E$, given by the sum of the kinetic $E_k$ and potential $E_p$ contributions, can be computed using potential theory in the limit of small amplitude oscillations, and results in an exponential decay of the form $E(t) = E_0 e^{-2\Gamma t}$, where $E_0$ is the total energy at time $t=0$ and $\Gamma= (8\pi^2\mu_w)/(\rho_w L^2)$. 
We approximate the total mechanical energy of the fluids with that of the water phase only ($\rho_w \gg \rho_a$), and write the kinetic and potential contributions as $E_k(t) = 0.5 \int_V \rho_w (\mathbf{u} \cdot  \mathbf{u}) \text{d}V$ and $E_p(t) = \int_V \rho_w g y \text{d}V - E_{p0}$; here, $V$ denotes the volume occupied by the water and $E_{p0} = - \rho_w g L (L/2)^2/2$ is the potential energy of the still water level. 
We thus compute the value of $E(t) = E_k(t) + E_p(t)$ throughout our simulations and plot it in panel $b$ of Fig.~\ref{fig:sketch}. 
All the curves are close to, but rather more dissipative than the analytical prediction \cite{landau-lifshitz-2003}, consistently with the experimental observations by \citet{luhar-nepf-2016}, who found that energy is dissipated slightly faster in the presence of a rigid stem, confirming that the stem rigidity plays a role in setting the rate of depletion of $E$. 
We nevertheless notice that, dissipation operates on a time scale significantly larger than the characteristic time of the stem motion.
The mechanical energy of the fluid at the final instant of our simulations ($t=5T$) is indeed $\approx20\%$ smaller than that at the beginning, corresponding to an attenuation of $\approx 10\%$ in the wave amplitude. 
The effect of this variation on the stem response is thus marginal, as discussed next in Fig. \ref{fig:tipMotion} of the results.

\section{Results}
\label{sec:results}

\subsection{The trajectories}

\begin{figure}
	\includegraphics[width=0.95\textwidth]{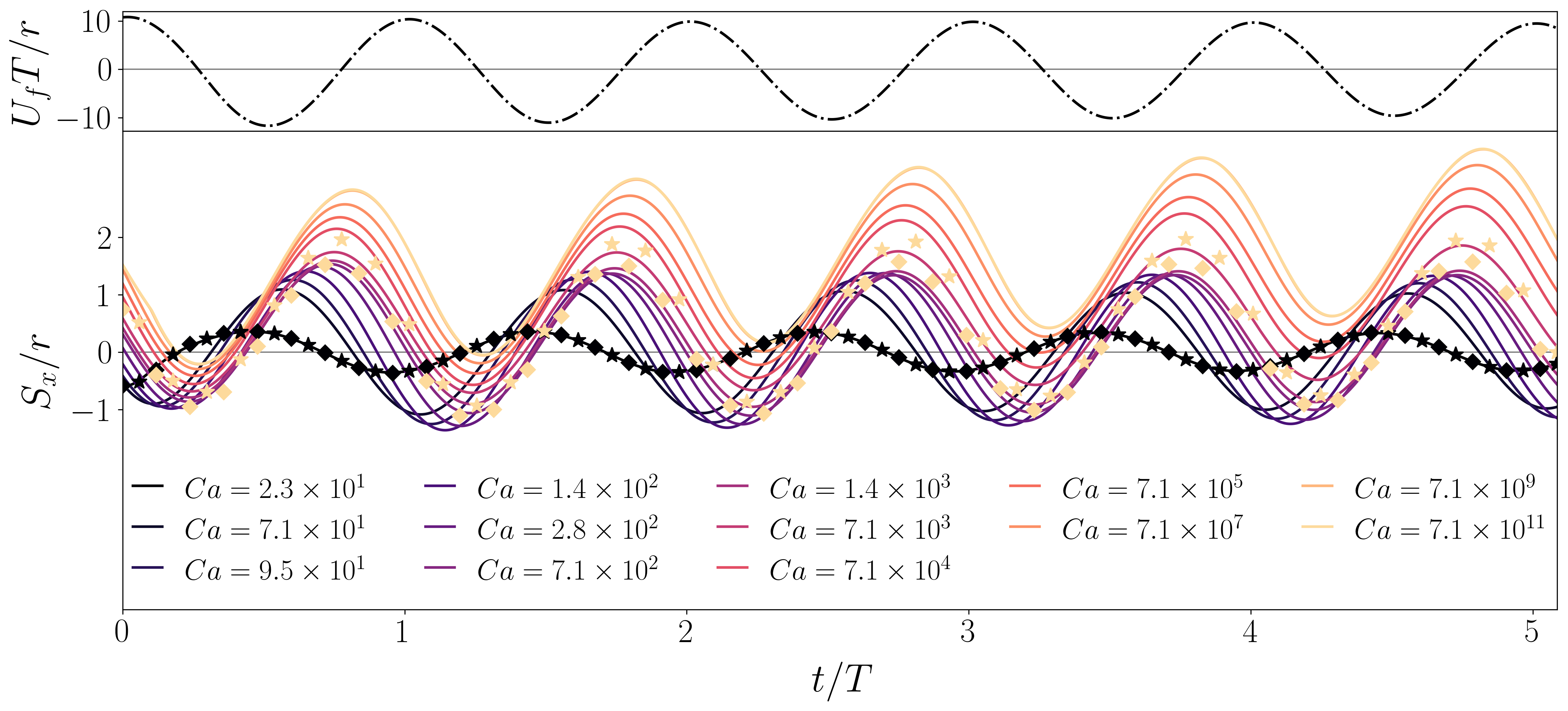}
	\put(-445,202){a)}
	\put(-445,162){b)}
	\caption{Motion of a flexible stem submerged under a decaying monochromatic gravity wave. In panel $a$ we show the streamwise velocity of the fluid measured at the undeflected stem tip position. In panel $b$, instead, we report the streamwise displacement of the stem tip from its undeflected position, varying the Cauchy number (colours) and the stem-to-water density ratio (symbols). Full lines refer to a density ratio of $0.95$, stars to a density ratio of $0.75$, and diamonds to a density ratio of $0.5$.}
	\label{fig:tipMotion}
\end{figure}

To investigate the motion of the stem, we start sampling the streamwise displacement of its tip from the undeflected configuration $S_x$ for $13$ different bending rigidity values; see Fig. \ref{fig:tipMotion}\textit{b}. The measurements are performed over more than $5$ wave periods after the release of the wave at $t=0$. For comparison, we also show in Fig. \ref{fig:tipMotion}\textit{a} the $x$ component of the unperturbed fluid velocity $U_f$ at the location of the undeflected stem tip (it is obtained from a simulation without the stem). Even thought the stem motion is induced by the integral effect of the flow field along its length and it is hindered by the clamp at the bottom wall, the displacement of the stem tip shares the same frequency of the wave ($f_{wave}=\omega/(2\pi)$) across all values of $Ca$, and exhibits an exponentially decaying sinusoidal behaviour in agreement with the damping of the wave. 
The curves for the two highest values of $Ca$ overlap almost perfectly (note that the difference is only barely visible in Fig. \ref{fig:tipMotion}\textit{b}), denoting the attainment of saturation in the case of an extremely flexible stem ($Ca \rightarrow \infty$).
For the lowest values of $Ca$ the stem fluctuates about its unperturbed configuration, and $S_x$ retains a null mean value. In contrast, for larger Cauchy numbers $S_x$ progressively deviates from a null mean value, compatibly with a downstream deflection of the stem in the longitudinal direction.
Notably, we observe that while at low $Ca$ the displacement $S_x$ is in phase-opposition with respect to the unperturbed velocity signal $U_f$, at higher $Ca$ it tends to synchronise with the fluctuations of the surrounding flow field.
As $Ca$ increases, indeed, the stem becomes more flexible and compliant with the fluid.
Upon reducing the density of the stem, we observe no significant variation of its behaviour in the most rigid case (the black line and symbols overlap almost exactly). 
In the most flexible one, instead, the stem exhibits smaller fluctuations, since it is more sensitive to the increased buoyant force that promotes its restoration to the straight and vertical configuration.

\begin{figure}
	\includegraphics[width=0.85\textwidth]{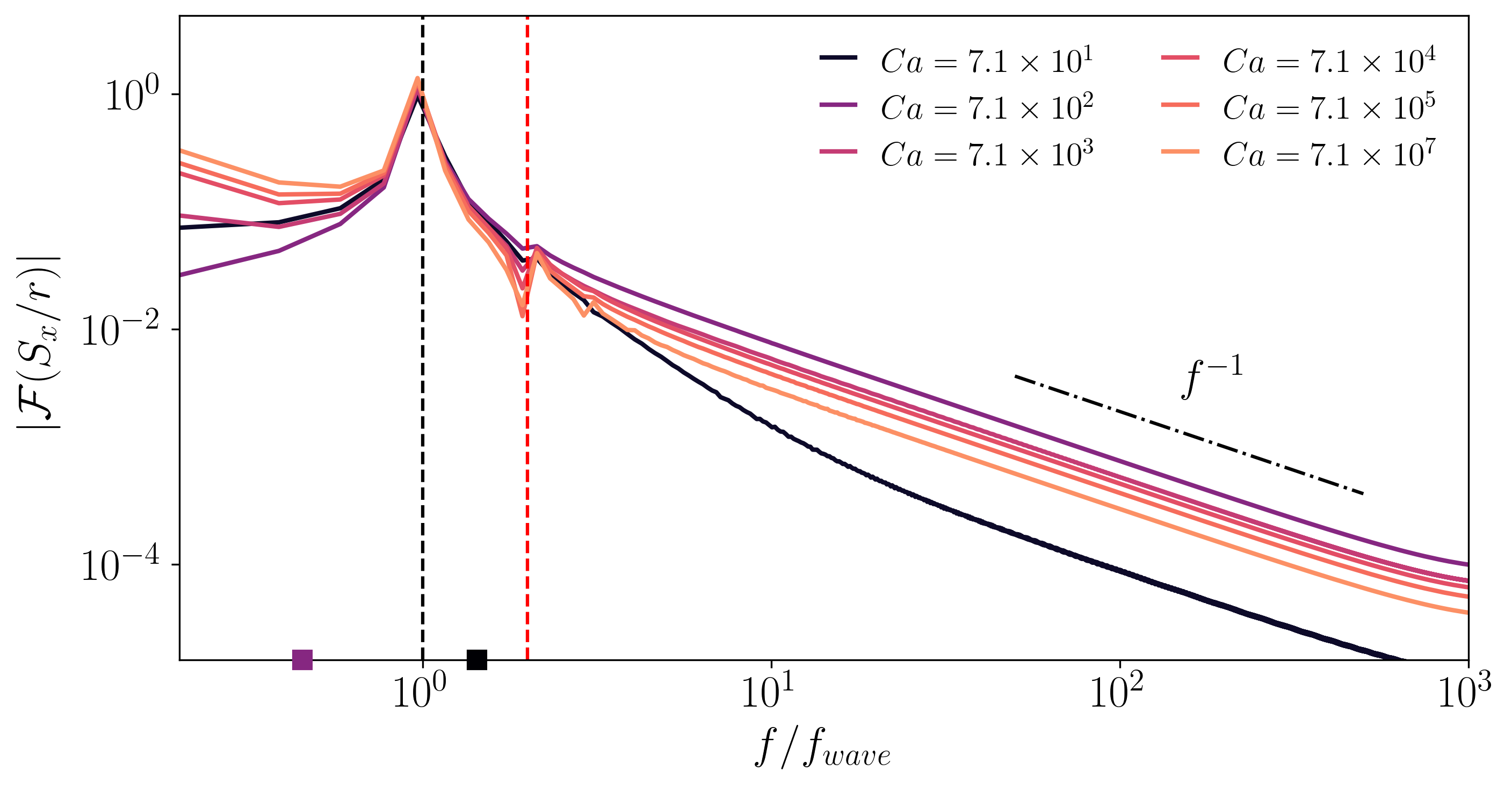}
	\caption{Temporal spectra of the stem tip displacement in the direction of propagation of the wave, for selected values of the Cauchy number. The wave frequency and its first harmonic are reported with a black and a red dashed line, respectively, while the black and purple squares on the x-axis correspond to the structural natural frequencies of the two most rigid stems, with all the others lying to the left outside the plot.}
	\label{fig:spectra}
\end{figure}

Since the wave decay is slow enough to only minor affect the fiber dynamics, we also compute the spectra of $S_x$ for selected values of $Ca$, shown in Fig. \ref{fig:spectra}. All of them exhibit a sharp peak in correspondence of the wave frequency (vertical black dashed line) and a minor one in correspondence of its first harmonic (vertical red dashed line), followed by a regular power-law decay $f^{-1}$ consistent with the arguments of \citet{jin-ji-chamorro-2016}. We also report the structural natural frequencies of the two most rigid stems (black and purple square on the x-axis, respectively), noticing that the spectra do not exhibit any deviation in their correspondence. The structural natural frequencies of all the other more flexible stems lie to the left of the reported range. We thus conclude that 
the stem response to the incoming wave occurs at the frequency of the wave itself, without any contribution at the structural natural frequency.

\begin{figure}
	\includegraphics[width=0.95\textwidth]{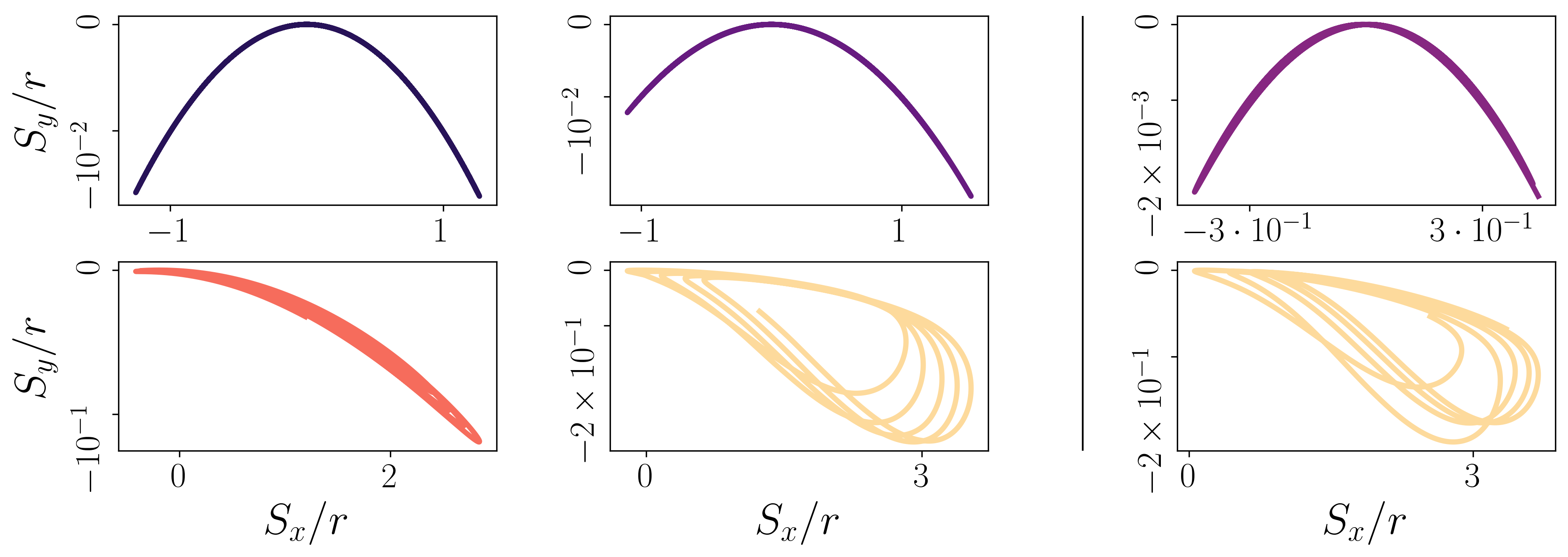}
	\put(-445,153){a)}
	\put(-293,153){b)}
	\put(-118,153){e)}	
	\put(-445,75){c)}
	\put(-293,75){d)}
	\put(-118,75){f)}
	
	\caption{Trajectory of the stem tip in the plane aligned with the longitudinal and vertical directions, passing through the clamp at the bottom wall. In panels $a-d$ we vary the Cauchy number, using a colour scale consistent with that adopted in Fig. \ref{fig:tipMotion}. Panels $e-f$, instead, report two additional validation cases. Panel $e$ corresponds to a stem with the same Cauchy number of panel $b$, but with a reduced length of $h/L = 0.2$. Panel $f$ reports the trajectory associated to the same stem of panel $d$, when placed under a wave released with a phase delay of $\pi$.}
	\label{fig:tipTrj}
\end{figure}
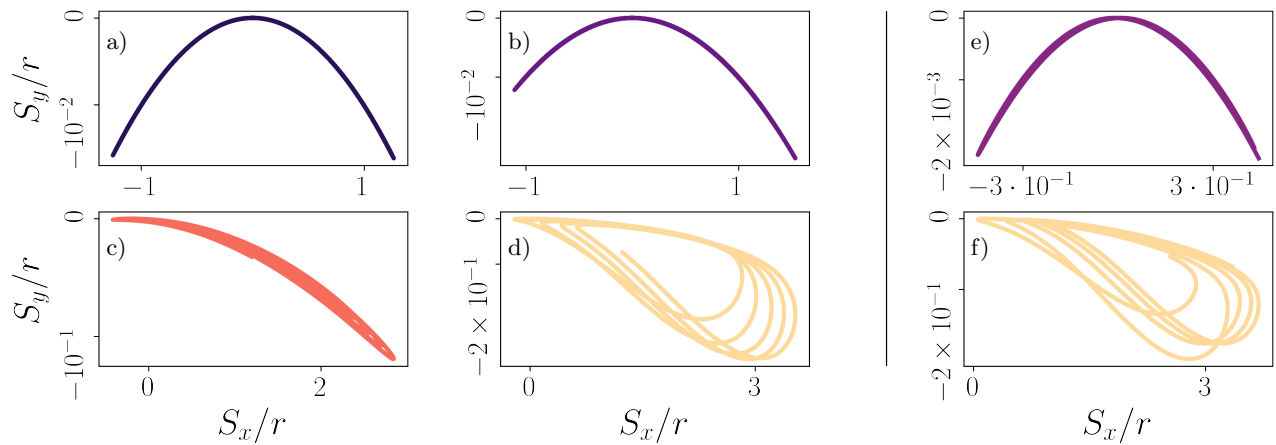

For all $Ca$ considered the stem exhibits a two-dimensional motion in the $x-y$ plane, with the displacement in the $z$ direction being negligible. 
We detail the dependence of the $x-y$ stem motion on the bending rigidity, and plot in Fig. \ref{fig:tipTrj} the tip displacement in the vertical direction $S_y$ against $S_x$ for four representative values of $Ca$. At the lowest values of $Ca$ ($Ca=95$, in panel $a$), the motion is substantially symmetric with respect to the vertical direction and the stem tip visits the same $(|S_x|,S_y)$ locations regardless of whether it is deflecting forward or backward. However, as $Ca$ is increased (in panels $b-d$) the symmetry is progressively broken and the stem exhibits a more accentuated deflection in the direction of wave propagation. In the most flexible case (panel $d$), all the points of the trajectory correspond to a positive deflection of the stem in the $x$ direction.
In addition, for the largest values of $Ca$ (in panels $c-d$) the trajectory opens and the stem tip visits different locations of the $S_x-S_y$ space according to the direction of its movement.
Our observations up to this point are consistent with former experimental studies on flexible blades clamped in gravity waves \cite{jacobsen-etal-2019} and associated dynamic models \cite{luhar-nepf-2016}. 

The loss of symmetry in the stem motion at the largest $Ca$ contrasts with the theory of ideal surface gravity waves \cite{lighthill-2001}, where the fluid particles describe periodic circular orbits under the wave and satisfy the symmetry with respect to the vertical direction. Yet, as reported in \S\ref{sec:setup_methods}, here we consider a finite-amplitude wave with $A=0.1$, responsible for the induction of a non-negligible Stokes drift \cite{longuett-stoneley-1953,bremer-breivik-2017}. The Stokes drift is the result of the average of the wave motion along a Lagrangian trajectory and predicts that a fluid particle, i.e. a tracer of negligible inertia, experiences a mean drift in the $x$ direction of propagation of the wave proportional to $U_0^2/c$, where $U_0 = A/\kappa$ is the amplitude of the wave induced velocity and $c$ is the wave phase velocity. Accordingly, for the largest $Ca$ the submerged stem is exposed to a finite mass flux and reconfigures \cite{luhar-nepf-2011}. This is particularly visible in Fig. \ref{fig:tipTrj}\textit{d}: flexible stems adapt to the motion of the surrounding fluid particles, which draw prolate trochoidal trajectories in the $S_x-S_y$ space. 

We have verified that the deflection of the stem is due to the Stokes drift only, being independent of the relative position (phase) of the stem under the wave at the release time $t=0$. Fig. \ref{fig:tipTrj}\textit{f} considers the same stem of panel $d$ ($Ca=7.1\times 10^{11}$), but placed with a phase delay of $\pi$ in the $x$ direction. The initial point of the trajectory clearly changes, but the stem shows the same downstream deflection as in panel \textit{d}. This confirms that the phase of the wave does not play a significant role in the reconfiguration of the stem, which is indeed probing the only symmetry breaking Lagrangian dynamics of the system induced by the Stokes drift. In Fig. \ref{fig:tipTrj}\textit{e}, instead, we address the dependence of the stem reconfiguration on $h$. We consider a stem with the same Cauchy number as in panel \textit{b} ($Ca=140$), but having a smaller length $h/L=0.2$ and a smaller rigidity to ensure no change in the Cauchy number. By doing so, we are able to isolate the dependence of the results from the length of the stem, and thus its submerged depth.
We keep the same resolution and discretise the stem with $45$ IBM Lagrangian points. As the stem lays deeper under the wave, it is exposed to a less intense fluid motion and it thus experiences smaller fluctuations. The fluid velocity field induced by the wave and the Stokes drift \cite{lighthill-2001} decays exponentially with the water depth; see Eqn. \eqref{eq:fluidFun}. We observe that for this value of the Cauchy number the drift velocity at $y/L=0.2$ is no more sufficient to induce the reconfiguration of the stem. 
The stem indeed responds to the fluid forcing without any time-averaged deflection, and simply oscillates about its undeflected configuration recovering a symmetric trajectory of its tip. Due to the interplay between the stem rigidity and the intensity of the fluid motion discussed above, two parameters ($Ca$ and $h/L$) are needed to exhaustively characterise the dynamics of the system at hand \cite{luhar-nepf-2016}. 
 
\subsection{The role of the natural frequency}

\begin{figure}
	\includegraphics[width=0.85\textwidth]{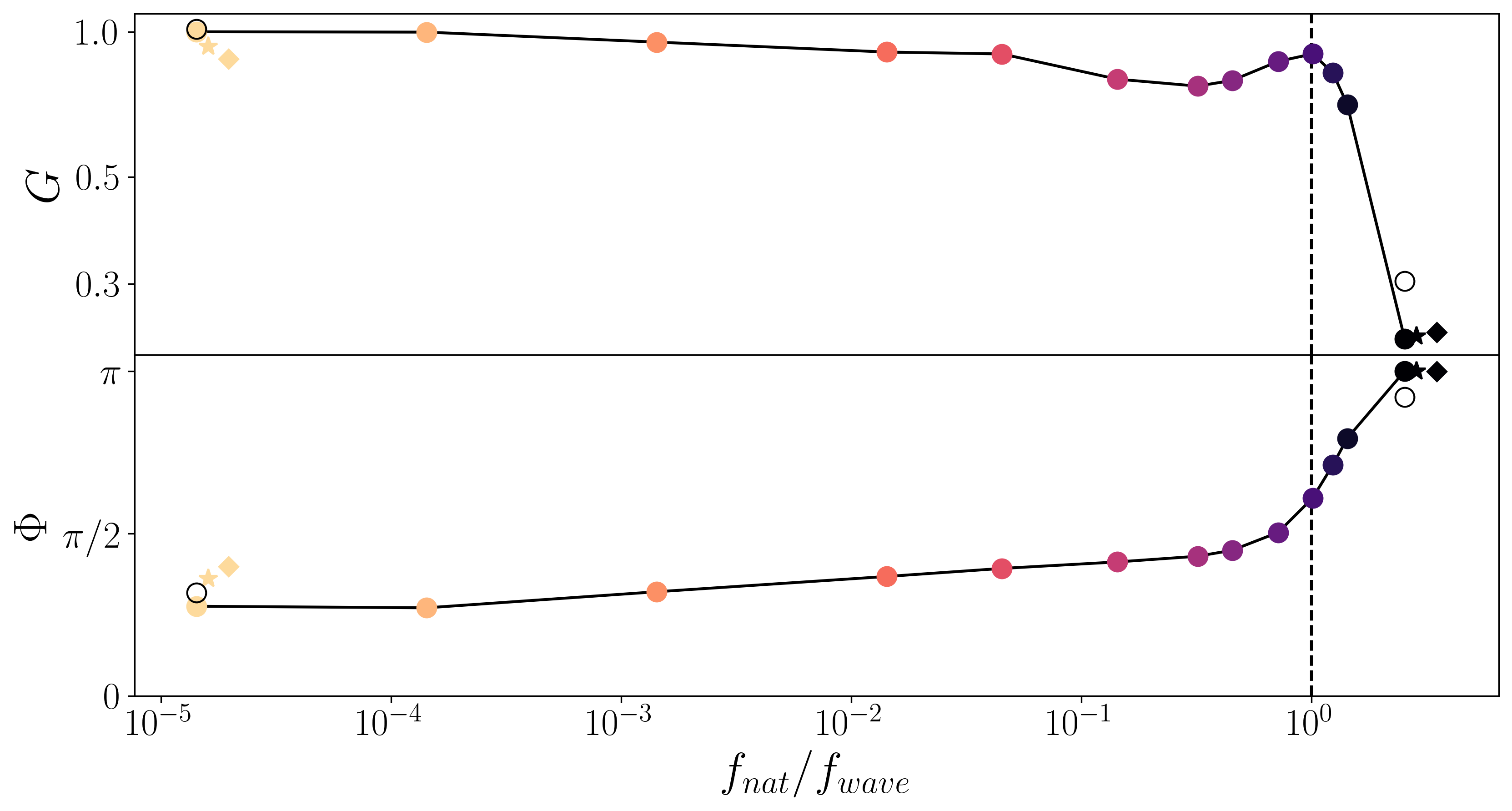}
	\put(-390,215){a)}
	\put(-390,117){b)}	
	\caption{Dynamical response of the stem as a function of the ratio between the structural natural frequency, $f_{nat}$, and the wave frequency, $f_{wave}$ (kept constant throughout the investigation). Panel $a$ measures the ratio $G$ between the peak-to-peak amplitude of the stem tip displacement in the longitudinal direction, and that attained for the highest value of $Ca$. Panel $b$, instead, quantifies the phase delay $\Phi$ of the stem tip displacement with respect to the fluid velocity at the undeflected stem tip location. In both panels, the circles correspond to the results with a stem-to-water density ratio of $0.95$, the stars to a density ratio of $0.75$, and the diamonds to a density ratio of $0.5$. Empty circles denote the results from two simulations at doubled grid resolution.}
	\label{fig:response}
\end{figure}

In the previous section we have observed that the rigidity of the stem (measured by the Cauchy number $Ca$) plays a key role in determining its response to the wave forcing, as it controls the interaction with the Stokes drift. The aim of this section is to delve into the stem dynamics in this set up, by highlighting how the natural dynamics of the stem influences the response. Note that in the simplified structural model adopted for this investigation, $Ca$ directly relates to the natural frequency of the stem $f_{nat}$ as
\begin{equation}
	f_{nat}\approx0.32\sqrt{\frac{1}{Ca}\frac{\rho_w}{\rho_s}\frac{Lg}{rh}}\sim Ca^{-0.5}.
\label{eq:fNat}
\end{equation}
To this end, in Fig. \ref{fig:response} we show the dynamical response of the stem as a function of the ratio between the structural natural frequency and the wave frequency $f_{nat}/f_{wave}$. Fig. \ref{fig:response}\textit{a} considers the gain of the response, represented by $G = \Delta S_{x}/\Delta S_{x}|_{Ca \rightarrow \infty}$, where $\Delta S = S_{x,max} - S_{x,min}$ is the amplitude of the tip oscillation estimated with the largest excursion the tip undergoes. $\Delta S_{x}|_{Ca \rightarrow \infty}$ is the amplitude of the $S_x$ fluctuations for the case in which the stem, despite clamped, is fully compliant to the flow; we actually use the value measured for $Ca = 7.1 \times 10^{11}$ (see Fig. \ref{fig:tipMotion}). Fig. \ref{fig:response}\textit{b}, instead, considers the phase delay $\Phi$ of the stem response with respect to the wave forcing. In particular, $\Phi$ is measured as the delay between the $S_x$ and the $U_f$ signals (see Fig. \ref{fig:tipMotion}), i.e. $\Phi = \measuredangle U_f - \measuredangle S_x$. Fig. \ref{fig:response} thus provides a simil  transfer function of the stem response to the wave forcing as a function of the $f_{nat}/f_{wave}$ frequency ratio. Note that, differently from what done usually in experiments \cite{kumar-etal-2021}, here we vary the structural rigidity of stem spanning a wide range of $f_{nat}$, while keeping the wave frequency fixed. Thus, while the representation in Fig. \ref{fig:response} resembles that commonly employed for transfer functions, here the abscissa is inversely proportional to the frequency of the forcing acting on the structure, which is the quantity kept constant throughout the investigation. 

Fig. \ref{fig:response}\textit{a} shows that $G$ plateaus about an almost unitary value when $f_{nat}/f_{wave}<1$, and drops abruptly as soon as $f_{nat}/f_{wave}>1$. This is consistent with Fig. \ref{fig:tipTrj} which shows that the $(S_x,S_y)$ fluctuations of the stem tip largely decrease as $Ca$ decreases.
Notably, we observe the presence of a (weak) peak in correspondence of $f_{nat}=f_{wave}$, denoting the occurrence of resonance when the natural frequency of the stem matches that of the wave. Such behaviour is a novelty for the drag-dominated oscillations investigated here ($KC\approx10$), where the natural dynamics of the structure is usually observed to be overcome by the wave dynamics \cite{mullarney-henderson-2010,kumar-etal-2021}. However, dynamical models for the stem motion \cite{kumar-etal-2021} typically account for linear waves only (i.e. small $A$), while the non-linear nature of the finite-amplitude wave we consider seems to be capable of triggering the natural dynamics of the stem.
It is also noticeable how the stem oscillations start getting significantly reduced compared to the most compliant case as soon as $f_{nat}/f_{wave}>1$, while for $f_{nat}/f_{wave}<1$ the effect is marginal. Clearly, $f_{nat}=f_{wave}$ represents a sharp cut-off for $G$ which marks a strong transition between two different dynamical regimes. 
Fig. \ref{fig:response}\textit{b} shows that, consistently with the observations of \citet{zhang-nakamura-2024}, the phase-delay between the wave and the stem motion varies with its structural rigidity and, thus, with the ratio $f_{nat}/f_{wave}$.
When $f_{nat}/f_{wave}\ll1$ the delay plateaus at a value of $\Phi \approx \pi/4$, sometimes also observed in oscillatory problems where the effect of viscous stresses is relatively small. Yet, here, we rather impute it to the finite deflection undergone by the flexible stem in the direction of propagation of the wave, and the consequent loss of symmetry.
The delay slightly increases with $f_{nat}/f_{wave}$, and reaches a value of $\Phi \approx \pi/2$ at resonance $f_{nat}/f_{wave} \approx 1$. Notably, the delay increases further for $f_{nat}/f_{wave}>1$, and the stem oscillations become in phase-opposition with $U_f$ in the most rigid case ($\Phi=\pi$). Compared to $G$, $\Phi$ exhibits a more regular behaviour and saturation is only attained for the two lowest values of $f_{nat}/f_{wave}$.
Next, we investigate how the density of the stem affects its response. A variation of the density (with the other structural parameters being fixed) determines a shift in the natural frequency of the stem, thus the displacement of the additional points in Fig. \ref{fig:response} along the horizontal direction. Yet, in terms of both magnitude and phase, the response is minimally affected in the most rigid case, consistently with what observed from Fig. \ref{fig:tipMotion}\textit{b}. In the most rigid case, instead, a smaller density density reduces the magnitude of the response and enhances its phase-lag, tending towards the behaviour of a more rigid stem at fixed density. This appears consistent with the enhancement of buoyancy.

To summarise, we highlight that the ratio $f_{nat}/f_{wave}$ determines the dynamical regime of the stem response to the wave. When $f_{nat}/f_{wave}\ll1$ large amplitude fluctuations occur with a modest lag on the fluid velocity, regardless of the actual value of $f_{nat}$. When $f_{nat}/f_{wave}\gg1$, instead, the fluctuations get increasingly smaller and are in phase-opposition with the fluid velocity. Resonance is found between the two regimes, when $f_{nat}/f_{wave} = 1$. Even though the stem oscillates at its natural frequency only in the peculiar resonant condition $f_{nat}=f_{wave}$, the natural dynamics of the stem plays an active role in determining the regime of its response to the wave forcing for all values of $f_{nat}/f_{wave}$. 

\begin{figure}
	\includegraphics[width=0.8\textwidth]{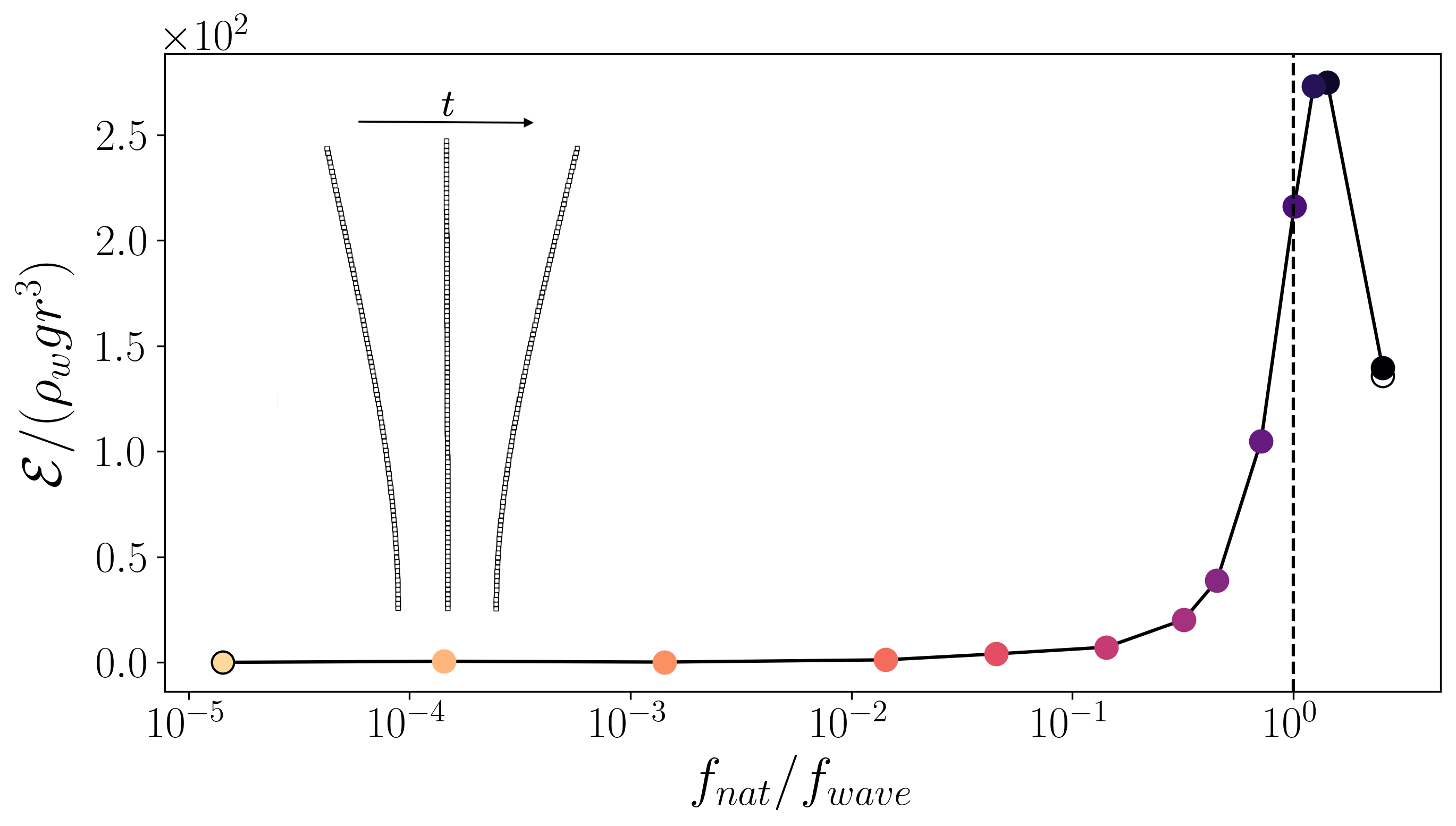}
	\caption{Elastic energy of the stem per unit length, averaged over one wave period. A sharp peak is observed close to the resonant condition $f_{nat}=f_{wave}$. Empty markers denote the results from two simulations at doubled grid resolution. We also show three side views (in the longitudinal-vertical plane) of the stem with the highest value of $\mathcal{E}$ ($Ca=95$), evenly sampled within one wave period. The axes of the stem are not in scale.}
	\label{fig:elEng}
\end{figure}

To better investigate the resonance found when $f_{nat}/f_{wave}=1$, we compute the elastic energy of the stem per unit length averaged over one wave period, i.e.
\begin{equation}
	\mathcal{E}=\frac{1}{{T}} \displaystyle\int_{{T}} \frac{1}{h} \displaystyle\int_h \gamma \frac{\partial^2 \mathbf{X}}{\partial s^2} \cdot \frac{\partial^2 \mathbf{X}}{\partial s^2} \mathrm{d}s \mathrm{d}t,
	\label{eq:elEng}
\end{equation}
and report its trend in Fig. \ref{fig:elEng}. We observe that $\mathcal{E}$ peaks close to $f_{nat}/f_{wave} = 1$, confirming the resonant nature of this state.
Yet, the actual value of $f_{nat}$ where $\mathcal{E}$ peaks is found slightly above the frequency of the forcing, as already observed in other dissipative systems \cite{foggirota-etal-2024-2}. 
For completeness, in the inset of Fig. \ref{fig:elEng} we show three side views of the stem with the highest value of $\mathcal{E}$ ($Ca=95$), evenly sampled within one wave period. We observe that the stem oscillates compatibly to its first structural mode, in agreement with what observed by \citet{kumar-etal-2021}. 

\section{Conclusions}
\label{sec:conclusions}

In this work, we have investigated the oscillations of a filamentous marine plant made by a single flexible stem under a monochromatic surface gravity wave of finite amplitude by means of fully resolved and coupled direct numerical simulations. 
The system at hand is composed by three distinct phases: air, water and structure. We chose realistic density and viscosity ratios for the two fluid phases, while spanning a wide range of rigidities for the structure. 
We have chosen proper values of the wave (in terms of length and amplitude) and stem (in terms of cross-sectional radius) parameters, to ensure that the motion of the stem occurs in the drag-dominated regime. 
The stem is always observed to sway at the same frequency of the wave ($f_{wave}$), yet different dynamical regimes emerge depending on the relative magnitude of the natural frequency of the stem $f_{nat}$ compared to $f_{wave}$. 
When $f_{nat}/f_{wave}\gg1$ the stem maintains, on average, an undeflected configuration, undergoing oscillations in phase-opposition with the fluid forcing and symmetric with respect to the vertical direction.
When $f_{nat}/f_{wave}\ll1$, instead, the stem reconfigures under the action of the Stokes drift induced by the finite-amplitude wave and bends forward, breaking the symmetry and moving more coherently with the flow field. 
At the transition between the two regimes (when $f_{nat}\approx f_{wave}$) resonance occurs, likely triggered by the non-linear nature of the wave, as confirmed by the trend of the elastic energy of the stem.

Consistently with previous observations in different setups \cite{foggirota-etal-2024}, our results confirm that also in this scenario the natural dynamics of the stem plays a fundamental role in determining the regime of its response to the forcing. However, in the present case the stem always oscillates at the frequency of the wave and fluctuations at the natural frequency only occur when $f_{nat}\approx f_{wave}$. 
This resonance within the drag-dominated regime was not detected in former experimental investigations \cite{mullarney-henderson-2010,kumar-etal-2021}, although also there the stem is observed to sway compatibly with its first structural mode.
While the numerical nature of our study allowed us to span a broader set of structural rigidities than what usually done in experiments, future investigations aimed at reconciling the two views should assess whether resonance can also be found in experiments, under an appropriate choice of parameters, or if instead the limited extent of the resonance reported here prevents its experimental identification. Additionally, further dissipation mechanisms not accounted for in our numerical schemes, e.g. the internal dissipation of the stem, might play a relevant role. \\

\section{Acknowledgements}
The research was supported by the Okinawa Institute of Science and Technology Graduate University (OIST) with subsidy funding to M.E.R. from the Cabinet Office, Government of Japan. 
M.E.R. also acknowledges funding from the Japan Society for the Promotion of Science (JSPS), grants \textit{24K17210} and \textit{24K00810}.
The authors acknowledge the computer time provided by the Scientific Computing section of the Core Facilities at OIST, and by the HPCI System Research Project with grants \textit{hp220402} and \textit{hp240006}.

\begin{figure}
	\includegraphics[width=0.9\textwidth]{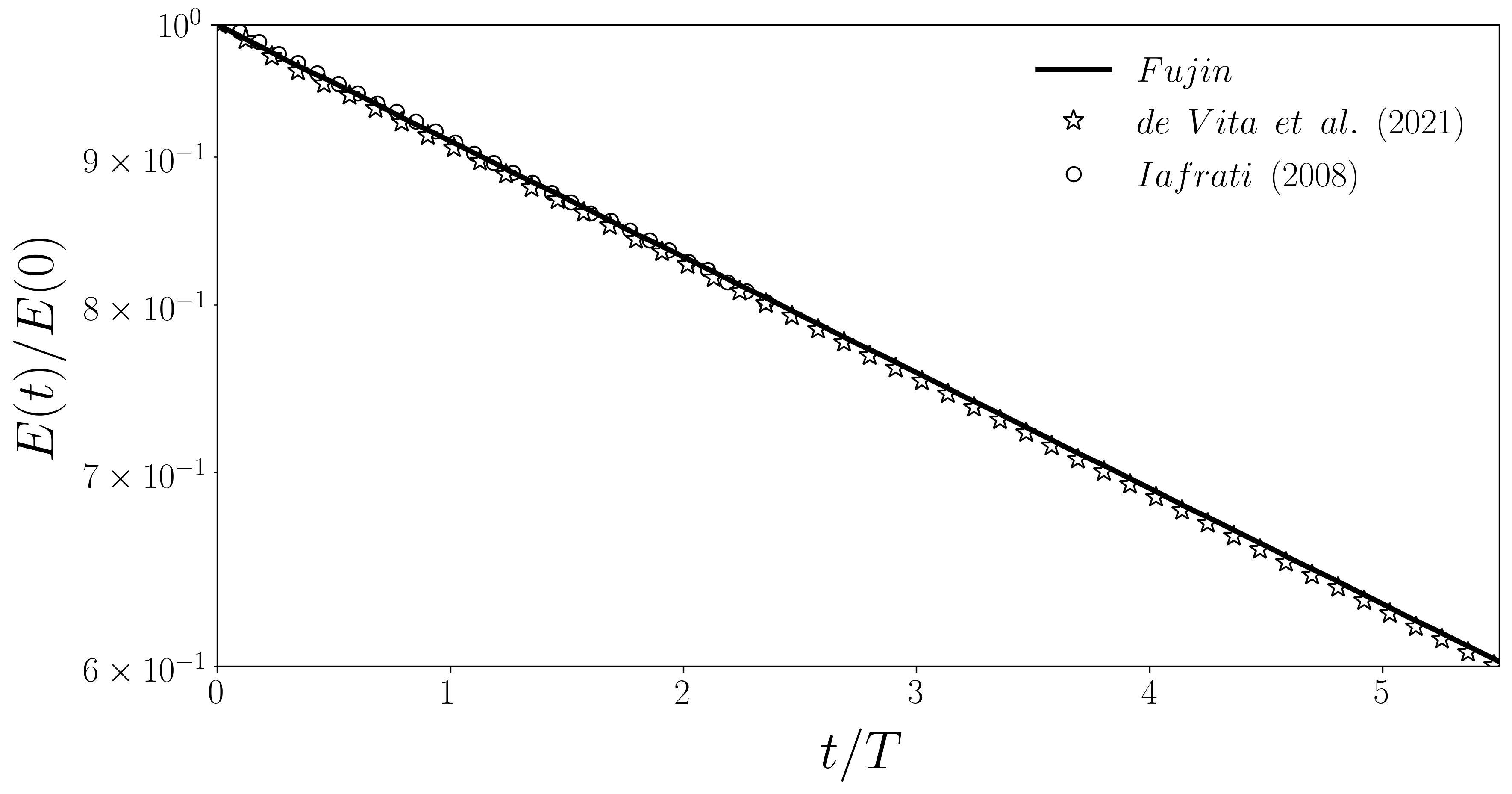}
	
	\caption{Code validation. We compare the decay of the mechanical energy of the wave, $E(t)$, predicted by our code (solid line) with those reported in literature \cite{iafrati-2009,devita-etal-2021} (symbols). A good matching is achieved.}
	\label{fig:validation}
\end{figure} 

\section{Validation}
\textit{Fujin}, the numerical solver employed for our investigation, has already been tested (\url{https://www.oist.jp/research/research-units/cffu/fujin}) and employed in multiple occurrences, for example to simulate a single-phase turbulent flow \cite{foggirota-etal-2023-2}, to investigate the coalescence of droplets dispersed in a carrier phase \cite{cannon-soligo-rosti-2024}, and to study the dynamics of flexible filaments and their interaction with the surrounding fluid \cite{monti-olivieri-rosti-2023}.
Relevant details on the validation of each component of the program (the core flow solver, the VOF, and the structural solver coupled to the fluid with a Lagrangian IBML method) are thus reported in literature. 
Here we further demonstrate the reliability of our numerical results by comparing the decay of the mechanical energy of the wave $E(t)$ predicted by our code with that reported by \citet{iafrati-2009} and \citet{devita-etal-2021}.
Consistently with \citet{devita-etal-2021}, we consider a case without the stem and initialise a wave with $A=0.05$, realistic density and viscosity ratios between the water and the air, and set the Reynolds number to $Re=10000$. Differently from what done in the cases where the stem is present, we impose the initial condition analytically, setting $t^*=0$ in Eqn.~\eqref{eq:fluidFun}, and immediately release the wave. The energy decay attained is reported in Fig. \ref{fig:validation} as a solid line, while stars correspond to the simulation of \citet{devita-etal-2021}: a close matching between the two is observed.
In the same figure we also report as circles the energy decay measured in a simulation by \citet{iafrati-2009}, over a shorter timespan. While in such case the amplitude of the wave is larger ($A=0.2$, potentially accentuating non-linear effects), the other parameters are unchanged and we still observe a remarkable correspondence.
We thus conclude that our solver is capable of accurately tackling the initialisation and decay of the monochromatic surface gravity wave considered in this investigation.


\begin{thebibliography}{52}%
\makeatletter
\providecommand \@ifxundefined [1]{%
 \@ifx{#1\undefined}
}%
\providecommand \@ifnum [1]{%
 \ifnum #1\expandafter \@firstoftwo
 \else \expandafter \@secondoftwo
 \fi
}%
\providecommand \@ifx [1]{%
 \ifx #1\expandafter \@firstoftwo
 \else \expandafter \@secondoftwo
 \fi
}%
\providecommand \natexlab [1]{#1}%
\providecommand \enquote  [1]{``#1''}%
\providecommand \bibnamefont  [1]{#1}%
\providecommand \bibfnamefont [1]{#1}%
\providecommand \citenamefont [1]{#1}%
\providecommand \href@noop [0]{\@secondoftwo}%
\providecommand \href [0]{\begingroup \@sanitize@url \@href}%
\providecommand \@href[1]{\@@startlink{#1}\@@href}%
\providecommand \@@href[1]{\endgroup#1\@@endlink}%
\providecommand \@sanitize@url [0]{\catcode `\\12\catcode `\$12\catcode
  `\&12\catcode `\#12\catcode `\^12\catcode `\_12\catcode `\%12\relax}%
\providecommand \@@startlink[1]{}%
\providecommand \@@endlink[0]{}%
\providecommand \url  [0]{\begingroup\@sanitize@url \@url }%
\providecommand \@url [1]{\endgroup\@href {#1}{\urlprefix }}%
\providecommand \urlprefix  [0]{URL }%
\providecommand \Eprint [0]{\href }%
\providecommand \doibase [0]{https://doi.org/}%
\providecommand \selectlanguage [0]{\@gobble}%
\providecommand \bibinfo  [0]{\@secondoftwo}%
\providecommand \bibfield  [0]{\@secondoftwo}%
\providecommand \translation [1]{[#1]}%
\providecommand \BibitemOpen [0]{}%
\providecommand \bibitemStop [0]{}%
\providecommand \bibitemNoStop [0]{.\EOS\space}%
\providecommand \EOS [0]{\spacefactor3000\relax}%
\providecommand \BibitemShut  [1]{\csname bibitem#1\endcsname}%
\let\auto@bib@innerbib\@empty
\bibitem [{\citenamefont {Alben}\ \emph {et~al.}(2002)\citenamefont {Alben},
  \citenamefont {Shelley},\ and\ \citenamefont
  {Zhang}}]{alben-shelley-zhang-2002}%
  \BibitemOpen
  \bibfield  {author} {\bibinfo {author} {\bibfnamefont {A.}~\bibnamefont
  {Alben}}, \bibinfo {author} {\bibfnamefont {M.}~\bibnamefont {Shelley}},\
  and\ \bibinfo {author} {\bibfnamefont {J.}~\bibnamefont {Zhang}},\ }\bibfield
   {title} {\bibinfo {title} {Drag reduction through self-similar bending of a
  flexible body},\ }\href {https://doi.org/10.1038/nature01232} {\bibfield
  {journal} {\bibinfo  {journal} {Nature}\ }\textbf {\bibinfo {volume} {420}},\
  \bibinfo {pages} {479} (\bibinfo {year} {2002})}\BibitemShut {NoStop}%
\bibitem [{\citenamefont {Gosselin}(2019)}]{gosselin-2019}%
  \BibitemOpen
  \bibfield  {author} {\bibinfo {author} {\bibfnamefont {F.}~\bibnamefont
  {Gosselin}},\ }\bibfield  {title} {\bibinfo {title} {Mechanics of a plant in
  fluid flow},\ }\href {https://doi.org/10.1093/jxb/erz288} {\bibfield
  {journal} {\bibinfo  {journal} {J. Exp. Bot.}\ }\textbf {\bibinfo {volume}
  {70}},\ \bibinfo {pages} {3533} (\bibinfo {year} {2019})}\BibitemShut
  {NoStop}%
\bibitem [{\citenamefont {Luhar}\ and\ \citenamefont
  {Nepf}(2011)}]{luhar-nepf-2011}%
  \BibitemOpen
  \bibfield  {author} {\bibinfo {author} {\bibfnamefont {M.}~\bibnamefont
  {Luhar}}\ and\ \bibinfo {author} {\bibfnamefont {H.~M.}\ \bibnamefont
  {Nepf}},\ }\bibfield  {title} {\bibinfo {title} {Flow-induced reconfiguration
  of buoyant and flexible aquatic vegetation},\ }\href
  {https://doi.org/10.4319/lo.2011.56.6.2003} {\bibfield  {journal} {\bibinfo
  {journal} {Limnol. Oceanogr.}\ }\textbf {\bibinfo {volume} {56}},\ \bibinfo
  {pages} {2003} (\bibinfo {year} {2011})}\BibitemShut {NoStop}%
\bibitem [{\citenamefont {Alben}(2008)}]{alben-2008}%
  \BibitemOpen
  \bibfield  {author} {\bibinfo {author} {\bibfnamefont {S.}~\bibnamefont
  {Alben}},\ }\bibfield  {title} {\bibinfo {title} {Optimal flexibility of a
  flapping appendage in an inviscid fluid},\ }\href
  {https://doi.org/10.1017/S0022112008003297} {\bibfield  {journal} {\bibinfo
  {journal} {J. Fluid Mech.}\ }\textbf {\bibinfo {volume} {614}},\ \bibinfo
  {pages} {355} (\bibinfo {year} {2008})}\BibitemShut {NoStop}%
\bibitem [{\citenamefont {Stanford}\ \emph {et~al.}(2008)\citenamefont
  {Stanford}, \citenamefont {Ifju}, \citenamefont {Albertani},\ and\
  \citenamefont {Shyy}}]{ifju-albertani-shyy-2008}%
  \BibitemOpen
  \bibfield  {author} {\bibinfo {author} {\bibfnamefont {B.}~\bibnamefont
  {Stanford}}, \bibinfo {author} {\bibfnamefont {P.}~\bibnamefont {Ifju}},
  \bibinfo {author} {\bibfnamefont {R.}~\bibnamefont {Albertani}},\ and\
  \bibinfo {author} {\bibfnamefont {W.}~\bibnamefont {Shyy}},\ }\bibfield
  {title} {\bibinfo {title} {Fixed membrane wings for micro air vehicles:
  {{Experimental}} characterization, numerical modeling, and tailoring},\
  }\href {https://doi.org/10.1016/j.paerosci.2008.03.001} {\bibfield  {journal}
  {\bibinfo  {journal} {Prog. Aerosp. Sci.}\ }\textbf {\bibinfo {volume}
  {44}},\ \bibinfo {pages} {258} (\bibinfo {year} {2008})}\BibitemShut
  {NoStop}%
\bibitem [{\citenamefont {Gosselin}\ \emph {et~al.}(2010)\citenamefont
  {Gosselin}, \citenamefont {Langre},\ and\ \citenamefont
  {{Machado-Almeida}}}]{gosselin-langre-machadoalmeida-2010}%
  \BibitemOpen
  \bibfield  {author} {\bibinfo {author} {\bibfnamefont {F.}~\bibnamefont
  {Gosselin}}, \bibinfo {author} {\bibfnamefont {E.}~\bibnamefont {Langre}},\
  and\ \bibinfo {author} {\bibfnamefont {B.~A.}\ \bibnamefont
  {{Machado-Almeida}}},\ }\bibfield  {title} {\bibinfo {title} {Drag reduction
  of flexible plates by reconfiguration},\ }\href
  {https://doi.org/10.1017/S0022112009993673} {\bibfield  {journal} {\bibinfo
  {journal} {J. Fluid Mech.}\ }\textbf {\bibinfo {volume} {650}},\ \bibinfo
  {pages} {319} (\bibinfo {year} {2010})}\BibitemShut {NoStop}%
\bibitem [{\citenamefont {Bearman}(1984)}]{bearman-1984}%
  \BibitemOpen
  \bibfield  {author} {\bibinfo {author} {\bibfnamefont {P.~W.}\ \bibnamefont
  {Bearman}},\ }\bibfield  {title} {\bibinfo {title} {Vortex {{Shedding}} from
  {{Oscillating Bluff Bodies}}},\ }\href
  {https://doi.org/10.1146/annurev.fl.16.010184.001211} {\bibfield  {journal}
  {\bibinfo  {journal} {Annu. Rev. Fluid Mech.}\ }\textbf {\bibinfo {volume}
  {16}},\ \bibinfo {pages} {195} (\bibinfo {year} {1984})}\BibitemShut
  {NoStop}%
\bibitem [{\citenamefont {Rosti}\ \emph {et~al.}(2018)\citenamefont {Rosti},
  \citenamefont {Banaei}, \citenamefont {Brandt},\ and\ \citenamefont
  {Mazzino}}]{rosti-etal-2018}%
  \BibitemOpen
  \bibfield  {author} {\bibinfo {author} {\bibfnamefont {M.~E.}\ \bibnamefont
  {Rosti}}, \bibinfo {author} {\bibfnamefont {A.~A.}\ \bibnamefont {Banaei}},
  \bibinfo {author} {\bibfnamefont {L.}~\bibnamefont {Brandt}},\ and\ \bibinfo
  {author} {\bibfnamefont {A.}~\bibnamefont {Mazzino}},\ }\bibfield  {title}
  {\bibinfo {title} {Flexible {{Fiber Reveals}} the {{Two-Point Statistical
  Properties}} of {{Turbulence}}},\ }\href
  {https://doi.org/10.1103/PhysRevLett.121.044501} {\bibfield  {journal}
  {\bibinfo  {journal} {Phys. Rev. Lett.}\ }\textbf {\bibinfo {volume} {121}},\
  \bibinfo {pages} {044501} (\bibinfo {year} {2018})}\BibitemShut {NoStop}%
\bibitem [{\citenamefont {Zhang}\ \emph {et~al.}(2000)\citenamefont {Zhang},
  \citenamefont {Childress}, \citenamefont {Libchaber},\ and\ \citenamefont
  {Shelley}}]{zhang-etal-2000}%
  \BibitemOpen
  \bibfield  {author} {\bibinfo {author} {\bibfnamefont {J.}~\bibnamefont
  {Zhang}}, \bibinfo {author} {\bibfnamefont {S.}~\bibnamefont {Childress}},
  \bibinfo {author} {\bibfnamefont {A.}~\bibnamefont {Libchaber}},\ and\
  \bibinfo {author} {\bibfnamefont {M.}~\bibnamefont {Shelley}},\ }\bibfield
  {title} {\bibinfo {title} {Flexible filaments in a flowing soap film as a
  model for one-dimensional flags in a two-dimensional wind},\ }\href
  {https://doi.org/10.1038/35048530} {\bibfield  {journal} {\bibinfo  {journal}
  {Nature}\ }\textbf {\bibinfo {volume} {408}},\ \bibinfo {pages} {835}
  (\bibinfo {year} {2000})}\BibitemShut {NoStop}%
\bibitem [{\citenamefont {Jung}\ \emph {et~al.}(2006)\citenamefont {Jung},
  \citenamefont {Mareck}, \citenamefont {Shelley},\ and\ \citenamefont
  {Zhang}}]{jung-etal-2006}%
  \BibitemOpen
  \bibfield  {author} {\bibinfo {author} {\bibfnamefont {S.}~\bibnamefont
  {Jung}}, \bibinfo {author} {\bibfnamefont {K.}~\bibnamefont {Mareck}},
  \bibinfo {author} {\bibfnamefont {M.}~\bibnamefont {Shelley}},\ and\ \bibinfo
  {author} {\bibfnamefont {J.}~\bibnamefont {Zhang}},\ }\bibfield  {title}
  {\bibinfo {title} {Dynamics of a {{Deformable Body}} in a {{Fast Flowing Soap
  Film}}},\ }\href {https://doi.org/10.1103/PhysRevLett.97.134502} {\bibfield
  {journal} {\bibinfo  {journal} {Phys. Rev. Lett.}\ }\textbf {\bibinfo
  {volume} {97}},\ \bibinfo {pages} {134502} (\bibinfo {year}
  {2006})}\BibitemShut {NoStop}%
\bibitem [{\citenamefont {Maza}\ \emph {et~al.}(2013)\citenamefont {Maza},
  \citenamefont {Lara},\ and\ \citenamefont {Losada}}]{maza-lara-losada-2013}%
  \BibitemOpen
  \bibfield  {author} {\bibinfo {author} {\bibfnamefont {M.}~\bibnamefont
  {Maza}}, \bibinfo {author} {\bibfnamefont {J.~L.}\ \bibnamefont {Lara}},\
  and\ \bibinfo {author} {\bibfnamefont {I.~J.}\ \bibnamefont {Losada}},\
  }\bibfield  {title} {\bibinfo {title} {A coupled model of submerged
  vegetation under oscillatory flow using {{Navier}}--{{Stokes}} equations},\
  }\href {https://doi.org/10.1016/j.coastaleng.2013.04.009} {\bibfield
  {journal} {\bibinfo  {journal} {Coast. Eng.}\ }\textbf {\bibinfo {volume}
  {80}},\ \bibinfo {pages} {16} (\bibinfo {year} {2013})}\BibitemShut {NoStop}%
\bibitem [{\citenamefont {Luhar}\ and\ \citenamefont
  {Nepf}(2016)}]{luhar-nepf-2016}%
  \BibitemOpen
  \bibfield  {author} {\bibinfo {author} {\bibfnamefont {M.}~\bibnamefont
  {Luhar}}\ and\ \bibinfo {author} {\bibfnamefont {H.~M.}\ \bibnamefont
  {Nepf}},\ }\bibfield  {title} {\bibinfo {title} {Wave-induced dynamics of
  flexible blades},\ }\href
  {https://doi.org/10.1016/j.jfluidstructs.2015.11.007} {\bibfield  {journal}
  {\bibinfo  {journal} {J. Fluids Struct.}\ }\textbf {\bibinfo {volume} {61}},\
  \bibinfo {pages} {20} (\bibinfo {year} {2016})}\BibitemShut {NoStop}%
\bibitem [{\citenamefont {Bagheri}\ \emph {et~al.}(2012)\citenamefont
  {Bagheri}, \citenamefont {Mazzino},\ and\ \citenamefont
  {Bottaro}}]{bagheri-mazzino-bottaro-2012}%
  \BibitemOpen
  \bibfield  {author} {\bibinfo {author} {\bibfnamefont {S.}~\bibnamefont
  {Bagheri}}, \bibinfo {author} {\bibfnamefont {A.}~\bibnamefont {Mazzino}},\
  and\ \bibinfo {author} {\bibfnamefont {A.}~\bibnamefont {Bottaro}},\
  }\bibfield  {title} {\bibinfo {title} {Spontaneous {{Symmetry Breaking}} of a
  {{Hinged Flapping Filament Generates Lift}}},\ }\href
  {https://doi.org/10.1103/PhysRevLett.109.154502} {\bibfield  {journal}
  {\bibinfo  {journal} {Phys. Rev. Lett.}\ }\textbf {\bibinfo {volume} {109}},\
  \bibinfo {pages} {154502} (\bibinfo {year} {2012})}\BibitemShut {NoStop}%
\bibitem [{\citenamefont {L{\=a}cis}\ \emph {et~al.}(2014)\citenamefont
  {L{\=a}cis}, \citenamefont {Brosse}, \citenamefont {Ingremeau}, \citenamefont
  {Mazzino}, \citenamefont {Lundell}, \citenamefont {Kellay},\ and\
  \citenamefont {Bagheri}}]{lacis-etal-2014}%
  \BibitemOpen
  \bibfield  {author} {\bibinfo {author} {\bibfnamefont {U.}~\bibnamefont
  {L{\=a}cis}}, \bibinfo {author} {\bibfnamefont {N.}~\bibnamefont {Brosse}},
  \bibinfo {author} {\bibfnamefont {F.}~\bibnamefont {Ingremeau}}, \bibinfo
  {author} {\bibfnamefont {A.}~\bibnamefont {Mazzino}}, \bibinfo {author}
  {\bibfnamefont {F.}~\bibnamefont {Lundell}}, \bibinfo {author} {\bibfnamefont
  {H.}~\bibnamefont {Kellay}},\ and\ \bibinfo {author} {\bibfnamefont
  {S.}~\bibnamefont {Bagheri}},\ }\bibfield  {title} {\bibinfo {title} {Passive
  appendages generate drift through symmetry breaking},\ }\href
  {https://doi.org/10.1038/ncomms6310} {\bibfield  {journal} {\bibinfo
  {journal} {Nat. Commun.}\ }\textbf {\bibinfo {volume} {5}},\ \bibinfo {pages}
  {5310} (\bibinfo {year} {2014})}\BibitemShut {NoStop}%
\bibitem [{\citenamefont {Jacobsen}\ \emph {et~al.}(2019)\citenamefont
  {Jacobsen}, \citenamefont {Bakker}, \citenamefont {Uijttewaal},\ and\
  \citenamefont {Uittenbogaard}}]{jacobsen-etal-2019}%
  \BibitemOpen
  \bibfield  {author} {\bibinfo {author} {\bibfnamefont {N.~G.}\ \bibnamefont
  {Jacobsen}}, \bibinfo {author} {\bibfnamefont {W.}~\bibnamefont {Bakker}},
  \bibinfo {author} {\bibfnamefont {W.~S.~J.}\ \bibnamefont {Uijttewaal}},\
  and\ \bibinfo {author} {\bibfnamefont {R.}~\bibnamefont {Uittenbogaard}},\
  }\bibfield  {title} {\bibinfo {title} {Experimental investigation of the
  wave-induced motion of and force distribution along a flexible stem},\ }\href
  {https://doi.org/10.1017/jfm.2019.739} {\bibfield  {journal} {\bibinfo
  {journal} {J. Fluid Mech.}\ }\textbf {\bibinfo {volume} {880}},\ \bibinfo
  {pages} {1036} (\bibinfo {year} {2019})}\BibitemShut {NoStop}%
\bibitem [{\citenamefont {Foggi~Rota}\ \emph
  {et~al.}(2024{\natexlab{a}})\citenamefont {Foggi~Rota}, \citenamefont
  {Koseki}, \citenamefont {Agrawal}, \citenamefont {Olivieri},\ and\
  \citenamefont {Rosti}}]{foggirota-etal-2024}%
  \BibitemOpen
  \bibfield  {author} {\bibinfo {author} {\bibfnamefont {G.}~\bibnamefont
  {Foggi~Rota}}, \bibinfo {author} {\bibfnamefont {M.}~\bibnamefont {Koseki}},
  \bibinfo {author} {\bibfnamefont {R.}~\bibnamefont {Agrawal}}, \bibinfo
  {author} {\bibfnamefont {S.}~\bibnamefont {Olivieri}},\ and\ \bibinfo
  {author} {\bibfnamefont {M.~E.}\ \bibnamefont {Rosti}},\ }\bibfield  {title}
  {\bibinfo {title} {Forced and natural dynamics of a clamped flexible fiber in
  wall turbulence},\ }\href {https://doi.org/10.1103/PhysRevFluids.9.L012601}
  {\bibfield  {journal} {\bibinfo  {journal} {Phys. Rev. Fluids}\ }\textbf
  {\bibinfo {volume} {9}},\ \bibinfo {pages} {L012601} (\bibinfo {year}
  {2024}{\natexlab{a}})}\BibitemShut {NoStop}%
\bibitem [{\citenamefont {Lighthill}(2001)}]{lighthill-2001}%
  \BibitemOpen
  \bibfield  {author} {\bibinfo {author} {\bibfnamefont {M.~J.}\ \bibnamefont
  {Lighthill}},\ }\href@noop {} {\emph {\bibinfo {title} {Waves in
  {{Fluids}}}}}\ (\bibinfo  {publisher} {Cambridge University Press},\ \bibinfo
  {year} {2001})\BibitemShut {NoStop}%
\bibitem [{\citenamefont {Landau}\ and\ \citenamefont
  {Lifshitz}(2003)}]{landau-lifshitz-2003}%
  \BibitemOpen
  \bibfield  {author} {\bibinfo {author} {\bibfnamefont {L.}~\bibnamefont
  {Landau}}\ and\ \bibinfo {author} {\bibfnamefont {E.}~\bibnamefont
  {Lifshitz}},\ }\href@noop {} {\emph {\bibinfo {title} {Fluid {{Mechanics}},
  {{Course}} of {{Theoretical Physics}}, {{Volume}} 6}}}\ (\bibinfo
  {publisher} {Butterworth-Heinmann},\ \bibinfo {year} {2003})\BibitemShut
  {NoStop}%
\bibitem [{\citenamefont {Chabchoub}\ \emph {et~al.}(2012)\citenamefont
  {Chabchoub}, \citenamefont {Akhmediev},\ and\ \citenamefont
  {Hoffmann}}]{chabchoub-akhmediev-hoffmann-2012}%
  \BibitemOpen
  \bibfield  {author} {\bibinfo {author} {\bibfnamefont {A.}~\bibnamefont
  {Chabchoub}}, \bibinfo {author} {\bibfnamefont {N.}~\bibnamefont
  {Akhmediev}},\ and\ \bibinfo {author} {\bibfnamefont {N.~P.}\ \bibnamefont
  {Hoffmann}},\ }\bibfield  {title} {\bibinfo {title} {Experimental study of
  spatiotemporally localized surface gravity water waves},\ }\href
  {https://doi.org/10.1103/PhysRevE.86.016311} {\bibfield  {journal} {\bibinfo
  {journal} {Phys. Rev. E}\ }\textbf {\bibinfo {volume} {86}},\ \bibinfo
  {pages} {016311} (\bibinfo {year} {2012})}\BibitemShut {NoStop}%
\bibitem [{\citenamefont {De~Vita}\ \emph {et~al.}(2018)\citenamefont
  {De~Vita}, \citenamefont {Verzicco},\ and\ \citenamefont
  {Iafrati}}]{devita-verzicco-iafrati-2018}%
  \BibitemOpen
  \bibfield  {author} {\bibinfo {author} {\bibfnamefont {F.}~\bibnamefont
  {De~Vita}}, \bibinfo {author} {\bibfnamefont {R.}~\bibnamefont {Verzicco}},\
  and\ \bibinfo {author} {\bibfnamefont {A.}~\bibnamefont {Iafrati}},\
  }\bibfield  {title} {\bibinfo {title} {Breaking of modulated wave groups:
  Kinematics and energy dissipation processes},\ }\href
  {https://doi.org/10.1017/jfm.2018.619} {\bibfield  {journal} {\bibinfo
  {journal} {J. Fluid Mech.}\ }\textbf {\bibinfo {volume} {855}},\ \bibinfo
  {pages} {267} (\bibinfo {year} {2018})}\BibitemShut {NoStop}%
\bibitem [{\citenamefont {De~Vita}\ \emph {et~al.}(2021)\citenamefont
  {De~Vita}, \citenamefont {De~Lillo}, \citenamefont {Bosia},\ and\
  \citenamefont {Onorato}}]{devita-etal-2021}%
  \BibitemOpen
  \bibfield  {author} {\bibinfo {author} {\bibfnamefont {F.}~\bibnamefont
  {De~Vita}}, \bibinfo {author} {\bibfnamefont {F.}~\bibnamefont {De~Lillo}},
  \bibinfo {author} {\bibfnamefont {F.}~\bibnamefont {Bosia}},\ and\ \bibinfo
  {author} {\bibfnamefont {M.}~\bibnamefont {Onorato}},\ }\bibfield  {title}
  {\bibinfo {title} {Attenuating surface gravity waves with mechanical
  metamaterials},\ }\href {https://doi.org/10.1063/5.0048613} {\bibfield
  {journal} {\bibinfo  {journal} {Phys. Fluids}\ }\textbf {\bibinfo {volume}
  {33}},\ \bibinfo {pages} {047113} (\bibinfo {year} {2021})}\BibitemShut
  {NoStop}%
\bibitem [{\citenamefont {Lorenzo}\ \emph {et~al.}(2023)\citenamefont
  {Lorenzo}, \citenamefont {Pezzutto}, \citenamefont {De~Lillo}, \citenamefont
  {Ventrella}, \citenamefont {De~Vita}, \citenamefont {Bosia},\ and\
  \citenamefont {Onorato}}]{lorenzo-etal-2023}%
  \BibitemOpen
  \bibfield  {author} {\bibinfo {author} {\bibfnamefont {M.}~\bibnamefont
  {Lorenzo}}, \bibinfo {author} {\bibfnamefont {P.}~\bibnamefont {Pezzutto}},
  \bibinfo {author} {\bibfnamefont {F.}~\bibnamefont {De~Lillo}}, \bibinfo
  {author} {\bibfnamefont {F.~M.}\ \bibnamefont {Ventrella}}, \bibinfo {author}
  {\bibfnamefont {F.}~\bibnamefont {De~Vita}}, \bibinfo {author} {\bibfnamefont
  {F.}~\bibnamefont {Bosia}},\ and\ \bibinfo {author} {\bibfnamefont
  {M.}~\bibnamefont {Onorato}},\ }\bibfield  {title} {\bibinfo {title}
  {Attenuating surface gravity waves with an array of submerged resonators: An
  experimental study},\ }\href {https://doi.org/10.1017/jfm.2023.741}
  {\bibfield  {journal} {\bibinfo  {journal} {J. Fluid Mech.}\ }\textbf
  {\bibinfo {volume} {973}},\ \bibinfo {pages} {A16} (\bibinfo {year}
  {2023})}\BibitemShut {NoStop}%
\bibitem [{\citenamefont {McWilliams}(2023)}]{mcwilliams-2023}%
  \BibitemOpen
  \bibfield  {author} {\bibinfo {author} {\bibfnamefont {J.~C.}\ \bibnamefont
  {McWilliams}},\ }\bibfield  {title} {\bibinfo {title} {Surface waves and
  currents in aquatic vegetation},\ }\href
  {https://doi.org/10.1017/jfm.2023.50} {\bibfield  {journal} {\bibinfo
  {journal} {J. Fluid Mech.}\ }\textbf {\bibinfo {volume} {958}},\ \bibinfo
  {pages} {A14} (\bibinfo {year} {2023})}\BibitemShut {NoStop}%
\bibitem [{\citenamefont {Leclercq}\ and\ \citenamefont
  {De~Langre}(2018)}]{leclercq-delangre-2018}%
  \BibitemOpen
  \bibfield  {author} {\bibinfo {author} {\bibfnamefont {T.}~\bibnamefont
  {Leclercq}}\ and\ \bibinfo {author} {\bibfnamefont {E.}~\bibnamefont
  {De~Langre}},\ }\bibfield  {title} {\bibinfo {title} {Reconfiguration of
  elastic blades in oscillatory flow},\ }\href
  {https://doi.org/10.1017/jfm.2017.910} {\bibfield  {journal} {\bibinfo
  {journal} {J. Fluid Mech.}\ }\textbf {\bibinfo {volume} {838}},\ \bibinfo
  {pages} {606} (\bibinfo {year} {2018})}\BibitemShut {NoStop}%
\bibitem [{\citenamefont {Mullarney}\ and\ \citenamefont
  {Henderson}(2010)}]{mullarney-henderson-2010}%
  \BibitemOpen
  \bibfield  {author} {\bibinfo {author} {\bibfnamefont {J.~C.}\ \bibnamefont
  {Mullarney}}\ and\ \bibinfo {author} {\bibfnamefont {S.~M.}\ \bibnamefont
  {Henderson}},\ }\bibfield  {title} {\bibinfo {title} {Wave-forced motion of
  submerged single-stem vegetation},\ }\bibfield  {journal} {\bibinfo
  {journal} {J. Geophys. Res. Oceans}\ }\textbf {\bibinfo {volume} {115}},\
  \href {https://doi.org/10.1029/2010JC006448} {10.1029/2010JC006448} (\bibinfo
  {year} {2010})\BibitemShut {NoStop}%
\bibitem [{\citenamefont {Kumar}\ \emph {et~al.}(2021)\citenamefont {Kumar},
  \citenamefont {Kumar}, \citenamefont {Deepu},\ and\ \citenamefont
  {Ramya}}]{kumar-etal-2021}%
  \BibitemOpen
  \bibfield  {author} {\bibinfo {author} {\bibfnamefont {K.}~\bibnamefont
  {Kumar}}, \bibinfo {author} {\bibfnamefont {V.}~\bibnamefont {Kumar}},
  \bibinfo {author} {\bibfnamefont {P.}~\bibnamefont {Deepu}},\ and\ \bibinfo
  {author} {\bibfnamefont {P.}~\bibnamefont {Ramya}},\ }\bibfield  {title}
  {\bibinfo {title} {Oscillations of a flexible filament under surface gravity
  waves},\ }\href {https://doi.org/10.1103/PhysRevFluids.6.114004} {\bibfield
  {journal} {\bibinfo  {journal} {Phys. Rev. Fluids}\ }\textbf {\bibinfo
  {volume} {6}},\ \bibinfo {pages} {114004} (\bibinfo {year}
  {2021})}\BibitemShut {NoStop}%
\bibitem [{\citenamefont {Stokes}(1847)}]{stokes-1847}%
  \BibitemOpen
  \bibfield  {author} {\bibinfo {author} {\bibfnamefont {G.~G.}\ \bibnamefont
  {Stokes}},\ }\bibfield  {title} {\bibinfo {title} {On the {{Theory}} of
  {{Oscillatory Waves}}},\ }\href@noop {} {\bibfield  {journal} {\bibinfo
  {journal} {Trans. Cambridge Philos.}\ }\textbf {\bibinfo {volume} {8}},\
  \bibinfo {pages} {441} (\bibinfo {year} {1847})}\BibitemShut {NoStop}%
\bibitem [{\citenamefont {{Longuet-Higgins}}(1953)}]{longuett-stoneley-1953}%
  \BibitemOpen
  \bibfield  {author} {\bibinfo {author} {\bibfnamefont {M.~S.}\ \bibnamefont
  {{Longuet-Higgins}}},\ }\bibfield  {title} {\bibinfo {title} {Mass transport
  in water waves},\ }\href {https://doi.org/10.1098/rsta.1953.0006} {\bibfield
  {journal} {\bibinfo  {journal} {Phil. Trans. R. Soc. A.}\ }\textbf {\bibinfo
  {volume} {245}},\ \bibinfo {pages} {535} (\bibinfo {year}
  {1953})}\BibitemShut {NoStop}%
\bibitem [{\citenamefont {Santamaria}\ \emph {et~al.}(2013)\citenamefont
  {Santamaria}, \citenamefont {Boffetta}, \citenamefont {Afonso}, \citenamefont
  {Mazzino}, \citenamefont {Onorato},\ and\ \citenamefont
  {Pugliese}}]{santamaria-etal-2013}%
  \BibitemOpen
  \bibfield  {author} {\bibinfo {author} {\bibfnamefont {F.}~\bibnamefont
  {Santamaria}}, \bibinfo {author} {\bibfnamefont {G.}~\bibnamefont
  {Boffetta}}, \bibinfo {author} {\bibfnamefont {M.~M.}\ \bibnamefont
  {Afonso}}, \bibinfo {author} {\bibfnamefont {A.}~\bibnamefont {Mazzino}},
  \bibinfo {author} {\bibfnamefont {M.}~\bibnamefont {Onorato}},\ and\ \bibinfo
  {author} {\bibfnamefont {D.}~\bibnamefont {Pugliese}},\ }\bibfield  {title}
  {\bibinfo {title} {Stokes drift for inertial particles transported by water
  waves},\ }\href {https://doi.org/10.1209/0295-5075/102/14003} {\bibfield
  {journal} {\bibinfo  {journal} {EPL}\ }\textbf {\bibinfo {volume} {102}},\
  \bibinfo {pages} {14003} (\bibinfo {year} {2013})}\BibitemShut {NoStop}%
\bibitem [{\citenamefont {{van den Bremer}}\ and\ \citenamefont
  {Breivik}(2017)}]{bremer-breivik-2017}%
  \BibitemOpen
  \bibfield  {author} {\bibinfo {author} {\bibfnamefont {T.~S.}\ \bibnamefont
  {{van den Bremer}}}\ and\ \bibinfo {author} {\bibfnamefont
  {{\O}.}~\bibnamefont {Breivik}},\ }\bibfield  {title} {\bibinfo {title}
  {Stokes drift},\ }\href {https://doi.org/10.1098/rsta.2017.0104} {\bibfield
  {journal} {\bibinfo  {journal} {Phil. Trans. R. Soc. A.}\ }\textbf {\bibinfo
  {volume} {376}},\ \bibinfo {pages} {20170104} (\bibinfo {year}
  {2017})}\BibitemShut {NoStop}%
\bibitem [{\citenamefont {Popinet}(2018)}]{popinet-2018}%
  \BibitemOpen
  \bibfield  {author} {\bibinfo {author} {\bibfnamefont {S.}~\bibnamefont
  {Popinet}},\ }\bibfield  {title} {\bibinfo {title} {Numerical {{Models}} of
  {{Surface Tension}}},\ }\href
  {https://doi.org/10.1146/annurev-fluid-122316-045034} {\bibfield  {journal}
  {\bibinfo  {journal} {Annu. Rev. Fluid Mech.}\ }\textbf {\bibinfo {volume}
  {50}},\ \bibinfo {pages} {49} (\bibinfo {year} {2018})}\BibitemShut {NoStop}%
\bibitem [{\citenamefont {Yu}(2005)}]{yu-2005}%
  \BibitemOpen
  \bibfield  {author} {\bibinfo {author} {\bibfnamefont {Z.}~\bibnamefont
  {Yu}},\ }\bibfield  {title} {\bibinfo {title} {A {{DLM}}/{{FD}} method for
  fluid/flexible-body interactions},\ }\href
  {https://doi.org/10.1016/j.jcp.2004.12.026} {\bibfield  {journal} {\bibinfo
  {journal} {J. Comput. Phys.}\ }\textbf {\bibinfo {volume} {207}},\ \bibinfo
  {pages} {1} (\bibinfo {year} {2005})}\BibitemShut {NoStop}%
\bibitem [{\citenamefont {Banaei}\ \emph {et~al.}(2020)\citenamefont {Banaei},
  \citenamefont {Rosti},\ and\ \citenamefont
  {Brandt}}]{banaei-rosti-brandt-2020}%
  \BibitemOpen
  \bibfield  {author} {\bibinfo {author} {\bibfnamefont {A.~A.}\ \bibnamefont
  {Banaei}}, \bibinfo {author} {\bibfnamefont {M.~E.}\ \bibnamefont {Rosti}},\
  and\ \bibinfo {author} {\bibfnamefont {L.}~\bibnamefont {Brandt}},\
  }\bibfield  {title} {\bibinfo {title} {Numerical study of filament
  suspensions at finite inertia},\ }\href
  {https://doi.org/10.1017/jfm.2019.794} {\bibfield  {journal} {\bibinfo
  {journal} {J. Fluid Mech.}\ }\textbf {\bibinfo {volume} {882}},\ \bibinfo
  {pages} {A5} (\bibinfo {year} {2020})}\BibitemShut {NoStop}%
\bibitem [{\citenamefont {{di Giorgio}}\ \emph {et~al.}(2022)\citenamefont {{di
  Giorgio}}, \citenamefont {Pirozzoli},\ and\ \citenamefont
  {Iafrati}}]{digiorgio-pirozzoli-iafrati-2022}%
  \BibitemOpen
  \bibfield  {author} {\bibinfo {author} {\bibfnamefont {S.}~\bibnamefont {{di
  Giorgio}}}, \bibinfo {author} {\bibfnamefont {S.}~\bibnamefont {Pirozzoli}},\
  and\ \bibinfo {author} {\bibfnamefont {A.}~\bibnamefont {Iafrati}},\
  }\bibfield  {title} {\bibinfo {title} {On coherent vortical structures in
  wave breaking},\ }\href {https://doi.org/10.1017/jfm.2022.674} {\bibfield
  {journal} {\bibinfo  {journal} {J. Fluid Mech.}\ }\textbf {\bibinfo {volume}
  {947}},\ \bibinfo {pages} {A44} (\bibinfo {year} {2022})}\BibitemShut
  {NoStop}%
\bibitem [{\citenamefont {Iafrati}(2009)}]{iafrati-2009}%
  \BibitemOpen
  \bibfield  {author} {\bibinfo {author} {\bibfnamefont {A.}~\bibnamefont
  {Iafrati}},\ }\bibfield  {title} {\bibinfo {title} {Numerical study of the
  effects of the breaking intensity on wave breaking flows},\ }\href
  {https://doi.org/10.1017/S0022112008005302} {\bibfield  {journal} {\bibinfo
  {journal} {J. Fluid Mech.}\ }\textbf {\bibinfo {volume} {622}},\ \bibinfo
  {pages} {371} (\bibinfo {year} {2009})}\BibitemShut {NoStop}%
\bibitem [{\citenamefont {Keulegan}\ and\ \citenamefont
  {Carpenter}(1956)}]{keulegan-carpenter-1956}%
  \BibitemOpen
  \bibfield  {author} {\bibinfo {author} {\bibfnamefont {G.~H.}\ \bibnamefont
  {Keulegan}}\ and\ \bibinfo {author} {\bibfnamefont {L.~H.}\ \bibnamefont
  {Carpenter}},\ }\href@noop {} {\emph {\bibinfo {title} {Forces on
  {{Cylinders}} and {{Plates}} in an {{Oscillating Fluid}}}}}\ (\bibinfo
  {publisher} {U.S. Department of Commerce, National Bureau of Standards},\
  \bibinfo {year} {1956})\BibitemShut {NoStop}%
\bibitem [{\citenamefont {Kim}\ and\ \citenamefont
  {Moin}(1985)}]{kim-moin-1985}%
  \BibitemOpen
  \bibfield  {author} {\bibinfo {author} {\bibfnamefont {J.}~\bibnamefont
  {Kim}}\ and\ \bibinfo {author} {\bibfnamefont {P.}~\bibnamefont {Moin}},\
  }\bibfield  {title} {\bibinfo {title} {Application of a fractional-step
  method to incompressible {{Navier-Stokes}} equations},\ }\href
  {https://doi.org/10.1016/0021-9991(85)90148-2} {\bibfield  {journal}
  {\bibinfo  {journal} {J. Comput. Phys.}\ }\textbf {\bibinfo {volume} {59}},\
  \bibinfo {pages} {308} (\bibinfo {year} {1985})}\BibitemShut {NoStop}%
\bibitem [{\citenamefont {Dorr}(1970)}]{dorr-1970}%
  \BibitemOpen
  \bibfield  {author} {\bibinfo {author} {\bibfnamefont {F.~W.}\ \bibnamefont
  {Dorr}},\ }\bibfield  {title} {\bibinfo {title} {The {{Direct Solution}} of
  the {{Discrete Poisson Equation}} on a {{Rectangle}}},\ }\href@noop {}
  {\bibfield  {journal} {\bibinfo  {journal} {SIAM Rev.}\ }\textbf {\bibinfo
  {volume} {12}},\ \bibinfo {pages} {248} (\bibinfo {year} {1970})},\ \Eprint
  {https://arxiv.org/abs/2029223} {2029223} \BibitemShut {NoStop}%
\bibitem [{\citenamefont {Peskin}(2002)}]{peskin-2002}%
  \BibitemOpen
  \bibfield  {author} {\bibinfo {author} {\bibfnamefont {C.~S.}\ \bibnamefont
  {Peskin}},\ }\bibfield  {title} {\bibinfo {title} {The immersed boundary
  method},\ }\href {https://doi.org/10.1017/S0962492902000077} {\bibfield
  {journal} {\bibinfo  {journal} {Acta Numer.}\ }\textbf {\bibinfo {volume}
  {11}},\ \bibinfo {pages} {479} (\bibinfo {year} {2002})}\BibitemShut
  {NoStop}%
\bibitem [{\citenamefont {Huang}\ \emph {et~al.}(2007)\citenamefont {Huang},
  \citenamefont {Shin},\ and\ \citenamefont {Sung}}]{huang-etal-2007}%
  \BibitemOpen
  \bibfield  {author} {\bibinfo {author} {\bibfnamefont {W.~X.}\ \bibnamefont
  {Huang}}, \bibinfo {author} {\bibfnamefont {S.~J.}\ \bibnamefont {Shin}},\
  and\ \bibinfo {author} {\bibfnamefont {H.~J.}\ \bibnamefont {Sung}},\
  }\bibfield  {title} {\bibinfo {title} {Simulation of flexible filaments in a
  uniform flow by the immersed boundary method},\ }\href
  {https://doi.org/10.1016/j.jcp.2007.07.002} {\bibfield  {journal} {\bibinfo
  {journal} {J. Comput. Phys.}\ }\textbf {\bibinfo {volume} {226}},\ \bibinfo
  {pages} {2206} (\bibinfo {year} {2007})}\BibitemShut {NoStop}%
\bibitem [{\citenamefont {Olivieri}\ \emph {et~al.}(2020)\citenamefont
  {Olivieri}, \citenamefont {Brandt}, \citenamefont {Rosti},\ and\
  \citenamefont {Mazzino}}]{olivieri-etal-2020-2}%
  \BibitemOpen
  \bibfield  {author} {\bibinfo {author} {\bibfnamefont {S.}~\bibnamefont
  {Olivieri}}, \bibinfo {author} {\bibfnamefont {L.}~\bibnamefont {Brandt}},
  \bibinfo {author} {\bibfnamefont {M.~E.}\ \bibnamefont {Rosti}},\ and\
  \bibinfo {author} {\bibfnamefont {A.}~\bibnamefont {Mazzino}},\ }\bibfield
  {title} {\bibinfo {title} {Dispersed {{Fibers Change}} the {{Classical Energy
  Budget}} of {{Turbulence}} via {{Nonlocal Transfer}}},\ }\href
  {https://doi.org/10.1103/PhysRevLett.125.114501} {\bibfield  {journal}
  {\bibinfo  {journal} {Phys. Rev. Lett.}\ }\textbf {\bibinfo {volume} {125}},\
  \bibinfo {pages} {114501} (\bibinfo {year} {2020})}\BibitemShut {NoStop}%
\bibitem [{\citenamefont {Monti}\ \emph {et~al.}(2023)\citenamefont {Monti},
  \citenamefont {Olivieri},\ and\ \citenamefont
  {Rosti}}]{monti-olivieri-rosti-2023}%
  \BibitemOpen
  \bibfield  {author} {\bibinfo {author} {\bibfnamefont {A.}~\bibnamefont
  {Monti}}, \bibinfo {author} {\bibfnamefont {S.}~\bibnamefont {Olivieri}},\
  and\ \bibinfo {author} {\bibfnamefont {M.~E.}\ \bibnamefont {Rosti}},\
  }\bibfield  {title} {\bibinfo {title} {Collective dynamics of dense hairy
  surfaces in turbulent flow},\ }\href
  {https://doi.org/10.1038/s41598-023-31534-7} {\bibfield  {journal} {\bibinfo
  {journal} {Sci. Rep.}\ }\textbf {\bibinfo {volume} {13}},\ \bibinfo {pages}
  {5184} (\bibinfo {year} {2023})}\BibitemShut {NoStop}%
\bibitem [{\citenamefont {Foggi~Rota}\ \emph
  {et~al.}(2024{\natexlab{b}})\citenamefont {Foggi~Rota}, \citenamefont
  {Monti}, \citenamefont {Olivieri},\ and\ \citenamefont
  {Rosti}}]{foggirota-etal-2024-2}%
  \BibitemOpen
  \bibfield  {author} {\bibinfo {author} {\bibfnamefont {G.}~\bibnamefont
  {Foggi~Rota}}, \bibinfo {author} {\bibfnamefont {A.}~\bibnamefont {Monti}},
  \bibinfo {author} {\bibfnamefont {S.}~\bibnamefont {Olivieri}},\ and\
  \bibinfo {author} {\bibfnamefont {M.~E.}\ \bibnamefont {Rosti}},\ }\bibfield
  {title} {\bibinfo {title} {Dynamics and fluid--structure interaction in
  turbulent flows within and above flexible canopies},\ }\href
  {https://doi.org/10.1017/jfm.2024.481} {\bibfield  {journal} {\bibinfo
  {journal} {J. Fluid Mech.}\ }\textbf {\bibinfo {volume} {989}},\ \bibinfo
  {pages} {A11} (\bibinfo {year} {2024}{\natexlab{b}})}\BibitemShut {NoStop}%
\bibitem [{\citenamefont {Hirt}\ and\ \citenamefont
  {Nichols}(1981)}]{hirt-nichols-1981}%
  \BibitemOpen
  \bibfield  {author} {\bibinfo {author} {\bibfnamefont {C.~W.}\ \bibnamefont
  {Hirt}}\ and\ \bibinfo {author} {\bibfnamefont {B.~D.}\ \bibnamefont
  {Nichols}},\ }\bibfield  {title} {\bibinfo {title} {Volume of fluid ({{VOF}})
  method for the dynamics of free boundaries},\ }\href
  {https://doi.org/10.1016/0021-9991(81)90145-5} {\bibfield  {journal}
  {\bibinfo  {journal} {J. Comp. Phys.}\ }\textbf {\bibinfo {volume} {39}},\
  \bibinfo {pages} {201} (\bibinfo {year} {1981})}\BibitemShut {NoStop}%
\bibitem [{\citenamefont {Quintard}\ and\ \citenamefont
  {Whitaker}(1994)}]{quintard-whitaker-1994a}%
  \BibitemOpen
  \bibfield  {author} {\bibinfo {author} {\bibfnamefont {M.}~\bibnamefont
  {Quintard}}\ and\ \bibinfo {author} {\bibfnamefont {S.}~\bibnamefont
  {Whitaker}},\ }\bibfield  {title} {\bibinfo {title} {Transport in ordered and
  disordered porous media {{I}}: {{The}} cellular average and the use of
  weighting functions},\ }\href@noop {} {\bibfield  {journal} {\bibinfo
  {journal} {Transport in Porous Med.}\ }\textbf {\bibinfo {volume} {14}},\
  \bibinfo {pages} {163} (\bibinfo {year} {1994})}\BibitemShut {NoStop}%
\bibitem [{\citenamefont {Ii}\ \emph {et~al.}(2012)\citenamefont {Ii},
  \citenamefont {Sugiyama}, \citenamefont {Takeuchi}, \citenamefont {Takagi},
  \citenamefont {Matsumoto},\ and\ \citenamefont {Xiao}}]{li-etal-2012}%
  \BibitemOpen
  \bibfield  {author} {\bibinfo {author} {\bibfnamefont {S.}~\bibnamefont
  {Ii}}, \bibinfo {author} {\bibfnamefont {K.}~\bibnamefont {Sugiyama}},
  \bibinfo {author} {\bibfnamefont {S.}~\bibnamefont {Takeuchi}}, \bibinfo
  {author} {\bibfnamefont {S.}~\bibnamefont {Takagi}}, \bibinfo {author}
  {\bibfnamefont {Y.}~\bibnamefont {Matsumoto}},\ and\ \bibinfo {author}
  {\bibfnamefont {F.}~\bibnamefont {Xiao}},\ }\bibfield  {title} {\bibinfo
  {title} {An interface capturing method with a continuous function: {{The
  THINC}} method with multi-dimensional reconstruction},\ }\href
  {https://doi.org/10.1016/j.jcp.2011.11.038} {\bibfield  {journal} {\bibinfo
  {journal} {J. Comp. Phys.}\ }\textbf {\bibinfo {volume} {231}},\ \bibinfo
  {pages} {2328} (\bibinfo {year} {2012})}\BibitemShut {NoStop}%
\bibitem [{\citenamefont {Rosti}\ \emph {et~al.}(2019)\citenamefont {Rosti},
  \citenamefont {De~Vita},\ and\ \citenamefont
  {Brandt}}]{rosti-devita-brandt-2019}%
  \BibitemOpen
  \bibfield  {author} {\bibinfo {author} {\bibfnamefont {M.~E.}\ \bibnamefont
  {Rosti}}, \bibinfo {author} {\bibfnamefont {F.}~\bibnamefont {De~Vita}},\
  and\ \bibinfo {author} {\bibfnamefont {L.}~\bibnamefont {Brandt}},\
  }\bibfield  {title} {\bibinfo {title} {Numerical simulations of emulsions in
  shear flows},\ }\href {https://doi.org/10.1007/s00707-018-2265-5} {\bibfield
  {journal} {\bibinfo  {journal} {Acta Mech.}\ }\textbf {\bibinfo {volume}
  {230}},\ \bibinfo {pages} {667} (\bibinfo {year} {2019})}\BibitemShut
  {NoStop}%
\bibitem [{\citenamefont {Hori}\ \emph {et~al.}(2023)\citenamefont {Hori},
  \citenamefont {Ng}, \citenamefont {Lohse},\ and\ \citenamefont
  {Verzicco}}]{hori-etal-2023}%
  \BibitemOpen
  \bibfield  {author} {\bibinfo {author} {\bibfnamefont {N.}~\bibnamefont
  {Hori}}, \bibinfo {author} {\bibfnamefont {C.~S.}\ \bibnamefont {Ng}},
  \bibinfo {author} {\bibfnamefont {D.}~\bibnamefont {Lohse}},\ and\ \bibinfo
  {author} {\bibfnamefont {R.}~\bibnamefont {Verzicco}},\ }\bibfield  {title}
  {\bibinfo {title} {Interfacial-dominated torque response in liquid--liquid
  {{Taylor}}--{{Couette}} flows},\ }\href {https://doi.org/10.1017/jfm.2023.29}
  {\bibfield  {journal} {\bibinfo  {journal} {J. Fluid Mech.}\ }\textbf
  {\bibinfo {volume} {956}},\ \bibinfo {pages} {A15} (\bibinfo {year}
  {2023})}\BibitemShut {NoStop}%
\bibitem [{\citenamefont {Cannon}\ \emph {et~al.}(2024)\citenamefont {Cannon},
  \citenamefont {Soligo},\ and\ \citenamefont
  {Rosti}}]{cannon-soligo-rosti-2024}%
  \BibitemOpen
  \bibfield  {author} {\bibinfo {author} {\bibfnamefont {I.}~\bibnamefont
  {Cannon}}, \bibinfo {author} {\bibfnamefont {G.}~\bibnamefont {Soligo}},\
  and\ \bibinfo {author} {\bibfnamefont {M.~E.}\ \bibnamefont {Rosti}},\
  }\bibfield  {title} {\bibinfo {title} {Morphology of clean and
  surfactant-laden droplets in homogeneous isotropic turbulence},\ }\href
  {https://doi.org/10.1017/jfm.2024.380} {\bibfield  {journal} {\bibinfo
  {journal} {J. Fluid Mech.}\ }\textbf {\bibinfo {volume} {987}},\ \bibinfo
  {pages} {A31} (\bibinfo {year} {2024})}\BibitemShut {NoStop}%
\bibitem [{\citenamefont {Jin}\ \emph {et~al.}(2016)\citenamefont {Jin},
  \citenamefont {Ji},\ and\ \citenamefont {Chamorro}}]{jin-ji-chamorro-2016}%
  \BibitemOpen
  \bibfield  {author} {\bibinfo {author} {\bibfnamefont {Y.}~\bibnamefont
  {Jin}}, \bibinfo {author} {\bibfnamefont {S.}~\bibnamefont {Ji}},\ and\
  \bibinfo {author} {\bibfnamefont {L.~P.}\ \bibnamefont {Chamorro}},\
  }\bibfield  {title} {\bibinfo {title} {Spectral energy cascade of body
  rotations and oscillations under turbulence},\ }\href
  {https://doi.org/10.1103/PhysRevE.94.063105} {\bibfield  {journal} {\bibinfo
  {journal} {Phys. Rev. E}\ }\textbf {\bibinfo {volume} {94}},\ \bibinfo
  {pages} {063105} (\bibinfo {year} {2016})}\BibitemShut {NoStop}%
\bibitem [{\citenamefont {Zhang}\ and\ \citenamefont
  {Nakamura}(2024)}]{zhang-nakamura-2024}%
  \BibitemOpen
  \bibfield  {author} {\bibinfo {author} {\bibfnamefont {J.}~\bibnamefont
  {Zhang}}\ and\ \bibinfo {author} {\bibfnamefont {T.}~\bibnamefont
  {Nakamura}},\ }\bibfield  {title} {\bibinfo {title} {Dynamics of a
  wall-mounted flexible plate in oscillatory flows},\ }\href
  {https://doi.org/10.1063/5.0214147} {\bibfield  {journal} {\bibinfo
  {journal} {Phys. Fluids}\ }\textbf {\bibinfo {volume} {36}},\ \bibinfo
  {pages} {073628} (\bibinfo {year} {2024})}\BibitemShut {NoStop}%
\bibitem [{\citenamefont {Foggi~Rota}\ \emph {et~al.}(2023)\citenamefont
  {Foggi~Rota}, \citenamefont {Monti}, \citenamefont {Rosti},\ and\
  \citenamefont {Quadrio}}]{foggirota-etal-2023-2}%
  \BibitemOpen
  \bibfield  {author} {\bibinfo {author} {\bibfnamefont {G.}~\bibnamefont
  {Foggi~Rota}}, \bibinfo {author} {\bibfnamefont {A.}~\bibnamefont {Monti}},
  \bibinfo {author} {\bibfnamefont {M.~E.}\ \bibnamefont {Rosti}},\ and\
  \bibinfo {author} {\bibfnamefont {M.}~\bibnamefont {Quadrio}},\ }\bibfield
  {title} {\bibinfo {title} {On--off pumping for drag reduction in a turbulent
  channel flow},\ }\href {https://doi.org/10.1017/jfm.2023.451} {\bibfield
  {journal} {\bibinfo  {journal} {J. Fluid Mech.}\ }\textbf {\bibinfo {volume}
  {966}},\ \bibinfo {pages} {A12} (\bibinfo {year} {2023})}\BibitemShut
  {NoStop}%
\end{thebibliography}
\end{document}